\documentclass[a4paper,10pt]{article}[2012/07/29]
\usepackage[cp1251]{inputenc}
\usepackage[english]{babel}
\usepackage{amsmath}
\usepackage{amssymb}
\usepackage{graphicx}
\usepackage[pdftex]{color}
\usepackage[sort&compress,square,numbers,comma]{natbib}
\usepackage[pdftex]{hyperref}
\hypersetup{
             final,
             colorlinks=true,
             linkcolor=blue,
             citecolor=red
}

\begin{document}
\twocolumn[\scriptsize{\slshape ISSN 1063-7761, Journal of Experimental and Theoretical Physics, 2012, Vol. 114, No. 3, pp. 511–527. \textcopyright\, Pleiades Publishing,
Inc., 2012.}

\scriptsize{\slshape Original Russian Text \textcopyright\, P.V. Ratnikov, A.P. Silin, 2012, published in Zhurnal \'{E}ksperimental’no\u{\i} i Teoretichesko\u{\i} Fiziki, 2012, Vol. 141, No. 3, pp. 582–601.}

\vspace{.75cm}

\rule{5.1cm}{.5pt}\hspace{6.8cm}\rule{5.1cm}{.5pt}

\vspace{-.17cm}

\rule{5.1cm}{.5pt}\hspace{6.8cm}\rule{5.1cm}{.5pt}

\begin{center}

\vspace{-1cm}
\large{\bf ELECTRONIC PROPERTIES}

\large{\bf OF SOLID}

\end{center}

\begin{center}
{\fontfamily{ptm}\fontsize{15.8pt}{20pt}\selectfont{\bf Size Quantization in Planar Graphene-Based Heterostructures:\\ \vspace{.25cm}Pseudospin Splitting, Interface States, and Excitons}}

\normalsize

\vspace{0.3cm}

\large{\bf P.\,V. Ratnikov$^*$ and A.\,P. Silin}

\vspace{0.1cm}

\normalsize

\textit{Lebedev Physical Institute, Russian Academy of Sciences,}

\textit{Leninski\u{\i} pr. 53, Moscow, 119991 Russia}

\textit{$^*$e-mail: \url{ratnikov@lpi.ru}}

Received April 21, 2011; in final form, August 1, 2011
\end{center}

\vspace{0.1cm}
\begin{list}{}
{\rightmargin=1cm \leftmargin=1cm}
\item
\small{{\bf Abstract}---A planar quantum-well device made of a gapless graphene nanoribbon with edges in contact with gapped graphene sheets is examined. The size-quantization spectrum of charge carriers in an asymmetric quantum well is shown to exhibit a pseudospin splitting. Interface states of a new type arise from the crossing of dispersion curves of gapless and gapped graphene materials. The exciton spectrum is calculated for a planar graphene quantum well. The effect of an external electric field on the exciton spectrum is analyzed.}

\vspace{0.05cm}

\small{\bf DOI}: 10.1134/S1063776112020094

\end{list}\vspace{0.5cm}]

\begin{center}
1. INTRODUCTION
\end{center}

The creation of graphene, a monolayer of carbon atoms forming a regular hexagonal lattice \citep{Novoselov1, Novoselov2, Zhang1}, has stimulated extensive experimental and theoretical studies along various lines of research. Graphene’s unique properties make it a promising material for a new generation of carbon-based nanoelectronic devices. In particular, carrier mobility in graphene amounts to $2\times10^5$ cm$^2$/V$\cdot$s, and ballistic transport is possible on a submicrometer scale \citep{Du, Morozov}.

Over the past seven years, numerous theoretical and experimental results have been reported on electronic properties of nanometer-wide ribbons of graphene (nanoribbons). Among the first were studies of electronic states of graphene nanoribbons using the Dirac equation under appropriate boundary conditions \citep{Brey1, Brey2}. The electronic properties of a graphene nanoribbon strongly depend on its size and edge geometry \citep{Son}. In terms of transport properties,
graphene nanoribbons are highly reminiscent of carbon nanotubes \citep{Saito, Ando1} since free carrier motion inside them is also one-dimensional.

A field-effect transistor (FET) based on a 2 nm wide and 236 nm long graphene nanoribbon was fabricated in a recent study \citep{Wang} (nanoribbons of widths between 10 and 60 nm were also studied). The graphene nanoribbon was made narrow enough to open a gap of width required for room-temperature transistor operation. However, it is less compact than the graphene quantum-dot transistor 30 nm in diameter discussed in \citep{Ponomarenko}.

In this study, we examine a planar quantum-well device made of a graphene nanoribbon whose edges are in contact with gapped graphene sheets.

A bandgap opening in graphene can be induced by several methods. First, graphene can be deposited on a hexagonal boron nitride (h--BN) substrate instead of a silicon-oxide one. This makes its two triangular sublattices nonequivalent, inducing in a bandgap of 53 meV \citep{Giovannetti}. Second, epitaxial graphene grown on a silicon-carbide substrate also has a nonzero bandgap \citep{Mattausch}. According to angle-resolved photoemission data, a bandgap of 0.26 eV is produced by this method \citep{Zhou}. Third, a hydrogenated derivative of graphene synthesized recently, graphane \citep{Elias}, has been predicted to have a direct bandgap of 5.4 eV at the $\Gamma$ point \citep{Lebegue}. Fourth, \textsl{ab initio} calculations have shown that CrO$_3$ adsorption on graphene induces a gap of 0.12 eV \cite{Zanella}. In the first two methods, a heterogeneous substrate can be used, such as an h--BN --- SiO$_2$ nanoribbon --- h--BN or SiC --- SiO$_2$ nanoribbon --- SiC one (\hyperlink{Fig1a}{Fig. 1a} depicts a substrate with h--BN). The last two methods produce a graphene sheet containing a nanoribbon without hydrogenation (as the nonhydrogenated one in \hyperlink{Fig1b}{Fig. 1b}) or a graphene strip without adsorbed CrO$_3$ molecules, respectively. Furthermore, the bandgap can be varied by using partially hydrogenated graphene (where some carbon atoms are not bonded to hydrogen atoms). Combinations of these methods can also be employed. Extensive experimental studies of graphene on substrates made of various materials, including rare-earth metals, have been reported recently \citep{Marchini, Martoccia, Pletikosic}. It may be possible to open a bandgap via adsorption of other molecules on graphene or by using other materials as substrates. The use of gapped graphene to create potential barriers opens up additional possibilities for bandgap engineering in carbon-based materials \citep{Han}.

We assume that both heterojunctions in the system combining a nanoribbon with gapped graphene sheets are type I junctions (e.g., see \citep{Vorobiev} for classification of junctions); i.e., the Dirac points in gapless graphene fall within the bandgaps of the adjacent gapped graphene sheets. This prevents spontaneous electron–hole pair creation, which would otherwise shunt graphene-based nanoelectronic devices such as FET.

We believe that planar heterostructures made of gapless and gapped graphene are as prospective building blocks in future carbon-based nanoelectronics. The use of only gapless graphene reduces the diverse opportunities offered by bandgap engineering in gapped graphene.

\begin{figure}[!t]
\hypertarget{Fig1a}{}
\includegraphics[width=0.5\textwidth]{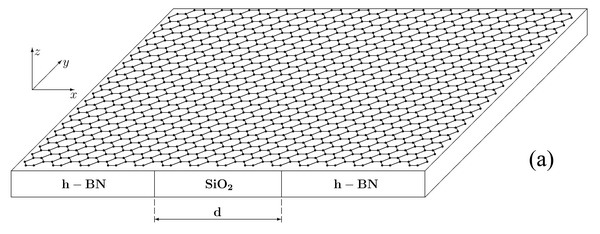}
\hypertarget{Fig1b}{}
\includegraphics[width=0.5\textwidth]{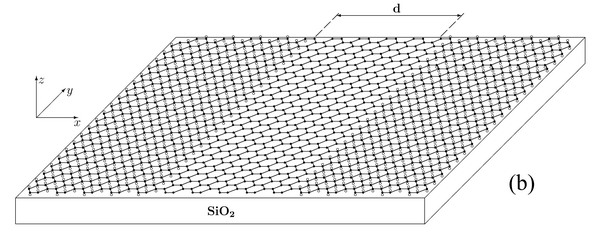}
{\bf Fig. 1.} Two configurations of the system under study: {\bf(a)}  graphene sheet on a substrate consisting of a SiO$_2$ nanoribbon of width d inserted between h--BN nanoribbons; {\bf(b)}
a graphene sheet on a SiO$_2$ substrate containing a nonhydrogenated nanoribbon of width  d,
where open and closed circles are hydrogen atoms bonded to carbon atoms in different sublattices
on opposite  sides of the sheet, respectively.
\end{figure}

The paper is organized as follows. The preliminary remarks in Section 2 recall some aspects of the theory of quasi-relativistic fermions in graphene. Section 3 introduces a generalized Dirac model for graphene. Section 4 describes the size quantization of originally massless carrier states in a planar graphene quantum well. Interface states are examined in Section 5. An effective bandgap opening in the size-quantization spectrum of the heterostructure in question makes it possible to generate excitons by optical pumping. The exciton spectrum is calculated in Section 6. In Section 7, the effect of an applied uniform electric field on an exciton is analyzed. Experimental manifestations of the effects examined in this paper are discussed in Section 8. The Conclusions section summarizes and discusses the main results obtained in this paper (Section 9).

\begin{center}
2. PRELIMINARY REMARKS
\end{center}

In our previous study \cite{Ratnikov1}, we analyzed a planar heterostructure combining two narrow-gap semiconductors with a graphene ribbon. We considered states with a definite parity $\lambda$, which is an eigenvalue of the parity operator \cite{Idlis}
\begin{equation}\label{1}
\widehat{P}=i\gamma_4\widehat{\Lambda}_{\bf n}.
\end{equation}
Here, $i\gamma_4$ is the inversion operator and
\begin{equation*}
\widehat{\Lambda}_{\bf n}=e^{-i\frac{\pi}{2}{\boldsymbol\Sigma}{\bf n}}=-i{\boldsymbol\Sigma}{\bf n}
\end{equation*}
is the operator of rotation by $\pi$ about an {\bf n}  axis perpendicular to the graphene plane. In standard representation,
\begin{equation*}
{\boldsymbol\Sigma}=\bigl(\begin{smallmatrix}
{\boldsymbol\sigma}&0\\0&{\boldsymbol\sigma}\end{smallmatrix}\bigr),
\end{equation*}
where ${\boldsymbol\sigma}$ denotes Pauli matrices, and
\begin{equation*}
\gamma_4\equiv\beta=\bigl(\begin{smallmatrix}I&0\\0&-I\end{smallmatrix}\bigr)
\end{equation*}
where $I$ is the 2$\times$2 unit matrix. It is clear that operator \eqref{1} is analogous to the parity operator $i\gamma_5\widehat{n}$ in quantum electrodynamics, where $\widehat{n}={\boldsymbol\gamma}{\bf
n}$,
\begin{equation*}
{\boldsymbol\gamma}=\bigl(\begin{smallmatrix}0&{\boldsymbol\sigma}\\-{\boldsymbol\sigma}&0\end{smallmatrix}\bigr),\hspace{.25cm}
\gamma_5=\gamma_1\gamma_2\gamma_3\gamma_4=i\bigl(\begin{smallmatrix}0&I\\I&0\end{smallmatrix}\bigr).
\end{equation*}
The eigenfunctions of this operator describe electron polarization states \cite{Akchiezer}.

\begin{figure}[!t]
\hypertarget{Fig2}{}
\includegraphics[width=0.5\textwidth]{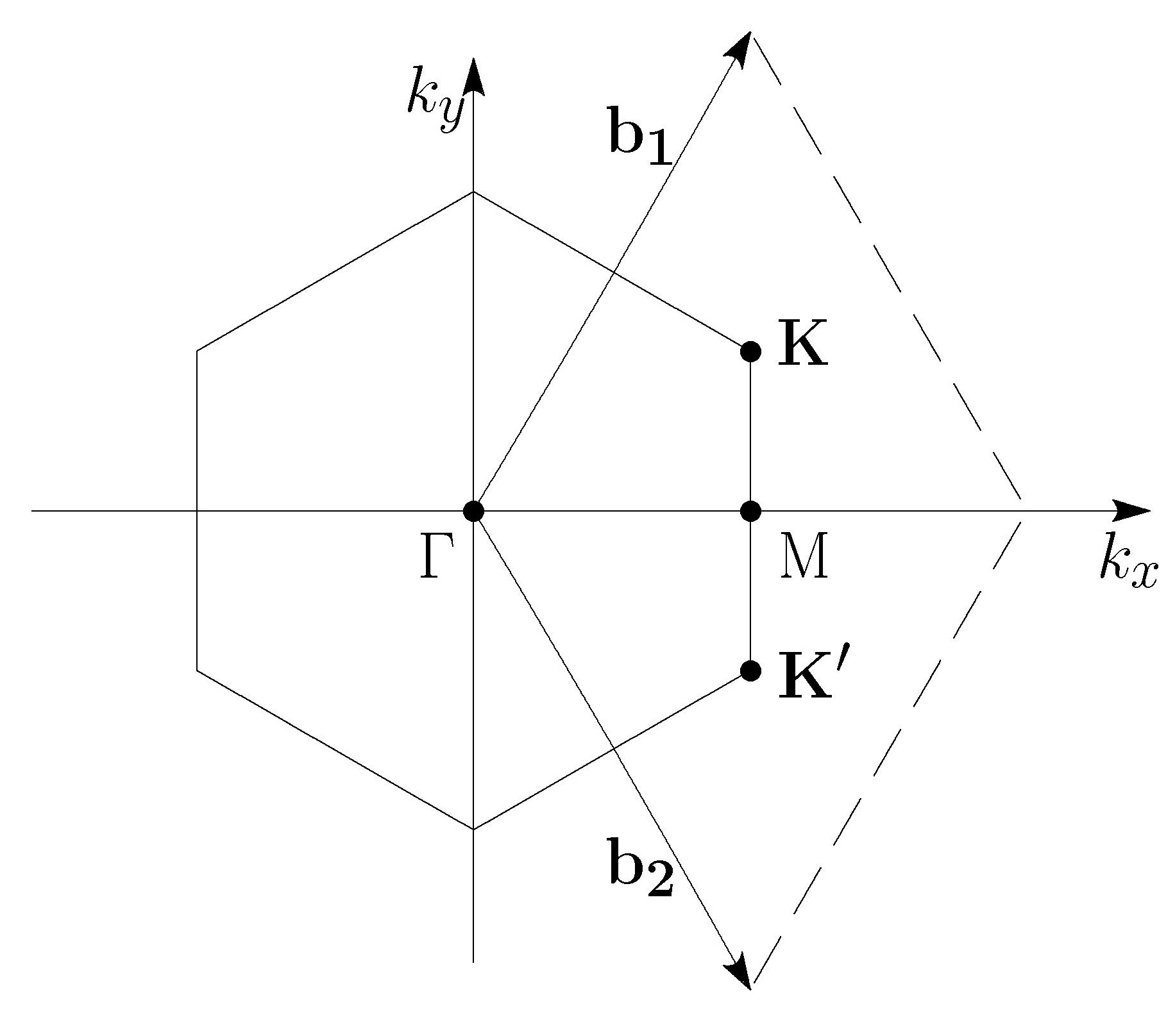}
{\bf Fig. 2.} First Brillouin zone of graphene, with linear energy spectrum at the corners (Dirac  points). The reciprocal lattice vectors {\bf b}$_1$ = $(2\pi/3a,\,2\pi/\sqrt{3}a)$ and  {\bf b}$_2$ = $(2\pi/3a,\,-2\pi/\sqrt{3}a)$, where  $a$  = 1.42 \AA\, is the lattice spacing, combined with the dashed lines equivalently represent the Brillouin zone as a rhombus.
\end{figure}

Charge carrier states in graphene can be described in terms of helicity define d as the eigenvalue of the operator  $\widehat{h}={\boldsymbol\sigma}\cdot{\bf p}/(2|{\bf p}|)$. The projection of pseudospin on the direction of quasimomentum p indicates the valley in the Brillouin zone where electrons or holes belong ($K$ or  $K^\prime$ point in \hyperlink{Fig2}{Fig. 2}). Positive helicity corresponds to electrons and holes with wavevectors near the $K$ and $K^\prime$ points, respectively; negative helicity corresponds to electrons and holes with wavevectors near the  $K^\prime$ and $K$ points, respectively \cite{Neto1}.

Massless states with opposite helicities are decoupled \cite{Lozovik1}. In addition, charge carriers have chiral symmetry (helicity is conserved), and parity can be defined for both massless and massive carriers\footnote{Massless states can be characterized by two quantum numbers:
helicity and sign of energy or helicity and eigenvalue of the operator $i\gamma_5$ \cite{Schweber}. Parity is analogous to the eigenvalue of $i\gamma_5$.}. In other words, a higher symmetry of massless charge carriers implies the existence of an additional quantum number: helicity. Whereas parity distinguishes only between the valleys where carrier states belong ($\lambda$ = 1 and –1 for states close to the $K$ and  $K^\prime$ points, respectively), helicity differs between a particle (electron) and an antiparticle (hole). However, chiral symmetry is broken for massive charge carriers (i.e., helicity is not a good quantum number any longer). Carrier states in a planar heterostructure combining gapless and gapped graphene should be characterized by parity.

Recall that the Dirac equation describing massless carriers in graphene in terms of 4$\times$4 matrices is derived by assuming that they are spinless and have two valley degrees of freedom \cite{Ando1}. When analysis is restricted to charge carriers in one valley, the Dirac equation can be reduced to a 2$\times$2 matrix representation by Weyl’s equation for a massless fermion analogous to neutrino in two Euclidean dimensions. The carrier energy spectrum with a pseudospin splitting in a planar heterostructure combining gapless and gapped graphene cannot be correctly analyzed in the 2$\times$2 representation. For similar reasons, the representation of the Dirac algebra in terms of 2$\times$2 matrices is not sufficient for describing the chiral symmetry breaking in quantum electrodynamics in two Euclidean dimensions \cite{Appelquist}.

Using the two-dimensional 4$\times$4 Dirac equation to describe charge carriers in a graphene-based nanostructure, we can study pseudospin effects following an approach to narrow-gap semiconductor heterostructures based on the Dirac model \cite{Volkov}. This makes methods developed for solving problems in the spintronics of narrow-gap semiconductor heterostructures applicable to graphene-based ones \cite{Kolesnikov1, Kolesnikov2, Silin1, Andryushin1, Andryushin2, Andryushin3, Andryushin4, Ratnikov2}.

As applied to a planar graphene-based heterostructure, these methods can be used to manipulate valley occupation in graphene via control of pseudospin. A heterostructure of this kind can be used in valleytronics \cite{Rycerz, Tworzydlo, Akhmerov, Xiao, Zhang2, Carcia, Pereira}. As a manifestation of pseudospin effect on charge carriers, we examine the pseudospin splitting of size-quantization spectrum (see Section 4 below).

\begin{center}
3. MODEL
\end{center}

To describe size quantization in graphene-based heterostructures, an equation containing a mass term should be written for the envelope wavefunction. A bandgap opening in the energy spectrum of graphene results from the lack of symmetry between the two triangular sublattices of its hexagonal lattice. The corresponding tight-binding Hamiltonian taking into account nearest-neighbor hopping has the form \cite{Semenoff}
\begin{equation*}
\widehat{H}=-t\sum_{{\bf B}, i, \sigma}\left[a^\dagger_\sigma({\bf
B}+{\bf d}_i)b_\sigma({\bf B})+b^\dagger_\sigma({\bf
B})a_\sigma({\bf B}+{\bf d}_i)\right]
\end{equation*}
\begin{equation}\label{2}
+\Delta\sum_{{\bf B}, \sigma}\left[a^\dagger_\sigma({\bf B}+{\bf
d}_1)a_\sigma({\bf B}+{\bf d}_1)-b^\dagger_\sigma({\bf
B})b_\sigma({\bf B})\right],
\end{equation}
where $t$ $\approx$ 2.8 eV is the nearest-neighbor hopping energy; the sum runs over the position vectors ${\bf B}$ of all $B$ sublattice atoms; the vectors ${\bf d}_i$ ($i$  = 1, 2, 3) pointing from a $B$ sublattice atom to the three nearest-neighbor $A$ sublattice atoms are expressed in terms of the lattice spacing $a$ as
\begin{equation*}
{\bf d_1}=\left(\frac{1}{2}a, \, \frac{\sqrt{3}}{2}a\right), \hspace{0.25cm}{\bf d_2}=\left(\frac{1}{2}a, \, -\frac{\sqrt{3}}{2}a\right),
\end{equation*}
\begin{equation*}
{\bf d_3}=\left(-a, \, 0\right)
\end{equation*}
(see \hyperlink{Fig3}{Fig. 3}); $\sigma=\uparrow,\,\downarrow$ is the (pseudo)spin index; $a_\sigma$ ($a^\dagger_\sigma$) and $b_\sigma$ ($b^\dagger_\sigma$) are the annihilation (creation) operators of $A$ and $B$ sublattice electrons, respectively; and the parameter $\Delta$ quantifies the on-site energy difference between the two sublattices (setting $\Delta$ = 0 restores the symmetry between sublattices so that graphene becomes gapless, whereas nonzero $\Delta$ equals the half-gap width in gapped graphene as shown below).

\begin{figure}[!t]
\hypertarget{Fig3}{}
\begin{center}
\includegraphics[width=0.5\textwidth]{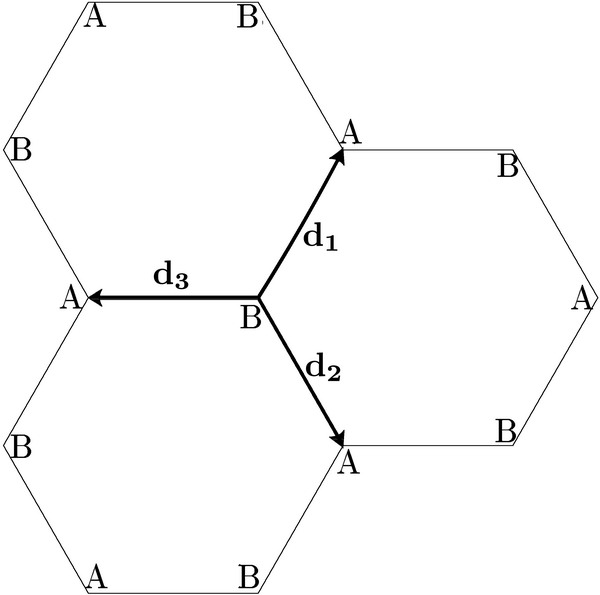}
\end{center}
{\bf Fig. 3.} Part of hexagonal lattice, with highlighted vectors {\bf d}$_i$ from a $B$ sublattice atom to the three nearest-neighbor $A$ sublattice atoms.
\end{figure}

Performing a Fourier transform, we change to the momentum representation
\begin{equation*}
a_\sigma({\bf A})=\int\limits_{\Omega_B}\frac{d^2k}{(2\pi)^2}a_\sigma({\bf k})e^{i{\bf k}\cdot{\bf A}},
\end{equation*}
\begin{equation*}
b_\sigma({\bf B})=\int\limits_{\Omega_B}\frac{d^2k}{(2\pi)^2}b_\sigma({\bf k})e^{i{\bf k}\cdot{\bf B}},
\end{equation*}
where $\Omega_B$ means integration over the first Brillouin zone.

Hamiltonian \eqref{2} is rewritten as
\begin{equation}\label{3}
\begin{split}
\widehat{H}=&\sum_\sigma\int\limits_{\Omega_B}\frac{d^2k}{(2\pi)^2}\begin{pmatrix} a^\dagger_\sigma({\bf k}) & b^\dagger_\sigma({\bf k})\end{pmatrix}\\ &\times\begin{pmatrix}\Delta & -t\sum\limits_ie^{-i{\bf k}\cdot{\bf d}_i}\\ -t\sum\limits_ie^{i{\bf k}\cdot{\bf d}_i} & -\Delta\end{pmatrix} \begin{pmatrix} a_\sigma({\bf k})\\ b_\sigma({\bf k})\end{pmatrix}.
\end{split}
\end{equation}
Conduction and valence band extrema lie at the corners of the Brillouin zone. We use Hamiltonian \eqref{3} expanded around the $K$ point with quasimomentum ${\bf q}_1=\left(\frac{2\pi}{3a},\,
\frac{2\pi}{3\sqrt{3}a}\right)$ or around the $K^\prime$ point with ${\bf q}_2=\left(\frac{2\pi}{3a},\, -\frac{2\pi}{3\sqrt{3}a}\right)$:
\begin{equation*}
\widehat{H}=\sum_\sigma\int\frac{d^2k}{(2\pi)^2}\widehat{\Psi}^\dagger_\sigma({\bf k})\widehat{\mathcal{H}}\widehat{\Psi}_\sigma({\bf k}).
\end{equation*}
where integration is performed over small neighbor-hoods of the $K$ and $K^\prime$ points. Near the corners, the Hamiltonian reduces to
\begin{equation}\label{4}
\widehat{\mathcal{H}}=\begin{pmatrix}v_F{\boldsymbol\sigma}\cdot{\bf k}+\Delta\sigma_z & 0\\ 0 & v_F{\boldsymbol\sigma}^*\cdot{\bf k}+\Delta\sigma_z\end{pmatrix},
\end{equation}
where $v_F=\frac{3}{2}at$ is the carrier Fermi velocity, ${\boldsymbol\sigma}=(\sigma_x,\, \sigma_y)$ and ${\boldsymbol\sigma}^*=(\sigma_x,\, -\sigma_y)$ are Pauli matrices in the
sublattice space, and $\widehat{\Psi}_\sigma({\bf k})$ is the bispinor defined as
\begin{equation*}
\widehat{\Psi}_\sigma({\bf
k})=\begin{pmatrix}\widehat{\Psi}^{(1)}_\sigma({\bf k})\\
\widehat{\Psi}^{(2)}_\sigma({\bf k})\end{pmatrix},
\end{equation*}
in terms of
\begin{equation*}
\widehat{\Psi}^{(1,2)}_\sigma({\bf k})=\exp\left(\frac{5\pi
i}{12}\sigma_z\right)\sigma_z\begin{pmatrix}a_\sigma({\bf q}_{1,2}+{\bf k})\\
b_\sigma({\bf q}_{1,2}+{\bf k})\end{pmatrix}.
\end{equation*}
We consider the heterostructure combining gapped gra-phene regions \textsl{1} ($x <$ 0) and \textsl{3} ($x >$ d) with a graphene nanoribbon \textsl{2} (0 $< x <$ d). The $z$ axis is perpendicular to the graphene plane, the $x$ axis is perpendicular to the heterojunction interfaces, and the $y$ axis is parallel to the interfaces (see \hyperlink{Fig1b}{Fig. 1}).

Using Hamiltonian \eqref{4}, we write an equation for the envelope wavefunction in a planar heterostructure:
\begin{equation}\label{5}
\begin{split}
&\left[v_{Fj}\left(\tau_0\otimes\sigma_x\widehat{p}_x+\tau_z\otimes\sigma_y\widehat{p}_y\right)\right.\\ &+\left.\tau_0\otimes\sigma_z\Delta_j+\tau_0\otimes\sigma_0\left(V_j-E\right)\right]\Psi(x,y)=0.
\end{split}
\end{equation}
Here, $\Delta_j=E_{gj}/2$ ($j=1,2,3$) denotes half-width of bandgap ($\Delta_1\neq0$ and $\Delta_3\neq0$ in regions \textsl{1} and \textsl{3}, whereas $\Delta_2=0$ region \textsl{2}); the respective work functions $V_1$ and $V_3$ of regions \textsl{1} and \textsl{3} depend on the mid-gap energies relative to the Dirac points for the corresponding materials (we set $V_2=0$ to be specific, see \hyperlink{Fig4}{Fig. 4}); the 2$\times$2
unit matrix $\sigma_0$ acts in the sublattice space; the 2$\times$2
unit matrix $\tau_0$ and the matrix $\tau_z$ defined similar to the Pauli matrix $\sigma_z$ act in the valley space; $\otimes$~is the Kronecker product symbol; and $\widehat{p}_x=-i\frac{\partial}{\partial x}$ and  $\widehat{p}_y=-i\frac{\partial}{\partial y}$ are the momentum operator components ($\hbar=1$). Assuming that the carrier Fermi velocities may differ between the three regions, we denote those for gapped regions \textsl{1}  and \textsl{3} by $v_{F1}$ and $v_{F3}$ and use $v_{F2}$ $\approx$ $10^8$ cm/s for gapless graphene.

Charge carriers move freely along the $y$ axis:
\begin{equation*}
\Psi(x,y)=\Psi(x)e^{ik_yy}.
\end{equation*}
The wavefunction $\Psi(x)$ is a bispinor:
\begin{equation*}
\Psi(x)=\begin{pmatrix}\psi_K(x)\\ \psi_{K^\prime}(x)\end{pmatrix},
\end{equation*}
where the spinors $\psi_K(x)$ and $\psi_{K^\prime}(x)$ represent charge carriers in the $K$ and $K^\prime$ valleys, respectively:
\begin{equation*}
\psi_K(x)=\begin{pmatrix}\psi_{KA}(x)\\
\psi_{KB}(x)\end{pmatrix},\hspace{0.2cm}\psi_{K^\prime}(x)=\begin{pmatrix}\psi_{K^\prime A}(x)\\
\psi_{K^\prime B}(x)\end{pmatrix}.
\end{equation*}
In the present context, the parity operator is expressed as follows:
\begin{equation}\label{6}
\widehat{P}=\tau_z\otimes\sigma_0.
\end{equation}

\begin{figure}[!t]
\hypertarget{Fig4}{}
\begin{center}
\includegraphics[width=0.5\textwidth]{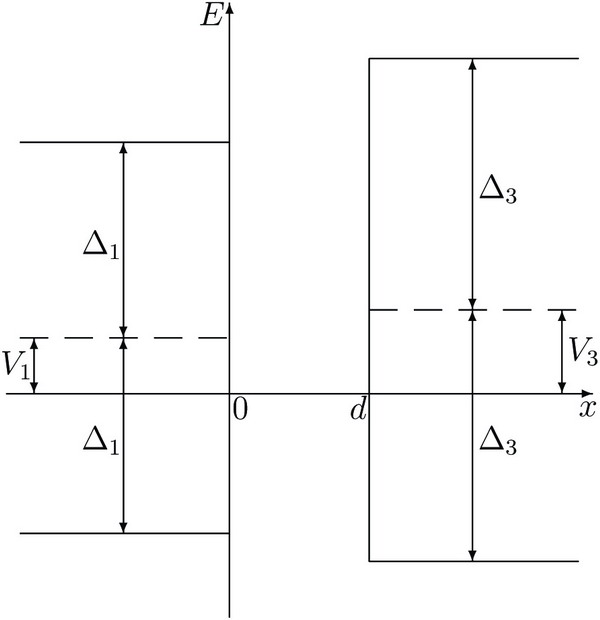}
\end{center}
{\bf Fig. 4.} The quantum well under analysis.
\end{figure}

Equation \eqref{5} is solved here in the parity basis. The eigenfunctions $\Psi_\lambda(x)$ of parity operator \eqref{6} are defined as follows:
\begin{equation*}
\widehat{P}\Psi_\lambda(x)=\lambda\Psi_\lambda(x),
\end{equation*}
\begin{equation}\label{7}
\begin{split}
\Psi_{+1}(x)&=\begin{pmatrix}\psi_{+1,K}(x)\\
0\end{pmatrix},\\ \Psi_{-1}(x)&=\begin{pmatrix}0\\
\psi_{-1,K^\prime}(x)\end{pmatrix}.
\end{split}
\end{equation}
Rewriting Eq. \eqref{5} as the 2$\times$2 matrix equations
\begin{equation}\label{8}
\begin{split}
&\left(-iv_{Fj}\sigma_x\frac{d}{dx}+v_{Fj}k_y\sigma_y+\lambda\Delta_j\sigma_z+V_j\right)\psi_{\lambda
K}(x)\\&=E_\lambda\psi_{\lambda K}(x),
\end{split}
\end{equation}
\begin{equation}\label{9}
\begin{split}
&\left(-iv_{Fj}\sigma_x\frac{d}{dx}-v_{Fj}k_y\sigma_y-\lambda\Delta_j\sigma_z+V_j\right)\psi_{\lambda
K^\prime}(x)\\&=E_\lambda\psi_{\lambda K^\prime}(x).
\end{split}
\end{equation}
We see that setting $\Delta_j=0$ and $V_j=0$ brings us back to the spinor wavefunctions describing chiral states near the $K$ or $K^\prime$ point, where the operator $\widehat{h}$ can be defined. However, chiral symmetry is broken when $\Delta\neq0$ (see previous section). Defining parity $\lambda$ as the eigenvalue of operator \eqref{6}, we find that it indicates the valley where charge carriers belong: by virtue of \eqref{7}, $\lambda$ = +1 for the states near the $K$ point described by Eq. \eqref{8} and $\lambda$ = –1 for the states near the $K^\prime$ point described by Eq.~\eqref{9}.

In both gapped and gapless graphene, the valleys transform into each other under time reversal. This is indicated by the opposite  signs of the terms proportional to $k_y$ in Eqs. \eqref{8} and \eqref{9}, since $k_y$ $\rightarrow$ $-k_y$ under time reversal. It can be shown directly by using the time reversal operator $\mathcal{T}$ in explicit form that  $\lambda$ $\rightarrow$ $-\lambda$ under $\mathcal{T}$. Indeed, if  $k_y$ is parallel to the line  $K-M-K^\prime$ (see \hyperlink{Fig2}{Fig. 2}) and its origin is set at $M$, then $K$ $\rightarrow$ $K^\prime$ and  $K^\prime$ $\rightarrow$ $K$ under $\mathcal{T}$.

Equations \eqref{8} and \eqref{9} are equivalently rewritten as the 2$\times$2 matrix equation
\begin{equation}\label{10}
\begin{split}
&\left(-iv_{Fj}\sigma_x\frac{d}{dx}+\lambda v_{Fj}k_y\sigma_y+\Delta_j\sigma_z+V_j\right)\psi_\lambda(x)\\&=E_\lambda\psi_\lambda(x).
\end{split}
\end{equation}
Hereinafter, valley indices $K$ and $K^\prime$ are omitted as unnecessary since $\lambda$ specifies the valley where charge carriers belong.

We now discuss the boundary conditions at the interfaces between different graphene materials. At the outset, we note that they are easier to formulate than those used at the graphene--free-space interface in models of edge states \cite{Brey1, Abanin}. To derive boundary conditions in the present model, we must find a relation between $\psi_\lambda(l)$ and $\psi_\lambda(-l)$ as $l$ $\rightarrow$ 0 in the neighbor-hood of x  = 0, where $l$ goes down to an atomic scale (condition at  x = d is derived similarly). Multiplying Eq. \eqref{10} by $\psi^\dagger(x)$ on the left, we integrate it over $[-l,\,l]$. Since  a is small, we neglect all terms except those containing a derivative with respect to $x$ to obtain\footnote{From the given equality, we have the continuity of the current component normal to the interface in the heterostructure plane. It is necessary condition.}
\begin{equation*}
\psi^{(-)\dagger}_\lambda(-l)v^{(-)}_F\psi^{(-)}_\lambda(-l)=\psi^{(+)\dagger}_\lambda(l)v^{(+)}_F\psi^{(+)}_\lambda(l),
\end{equation*}
where $\psi^{(-)}_\lambda$ and $\psi^{(+)}_\lambda$ are defined on the left- and right-hand sides of the boundary (at $x$ $<$ 0 and  $x$ $>$ 0, respectively). Representing these functions as
\begin{equation*}
\psi^{(\pm)}_\lambda=\left|\psi^{(\pm)}_\lambda\right|\exp\left(i\varphi^{(\pm)}\right),
\end{equation*}
we rewrite the equality above as
\begin{equation*}
\sqrt{v^{(-)}_F}\left|\psi^{(-)\dagger}_\lambda(-l)\right|=\sqrt{v^{(+)}_F}\left|\psi^{(+)\dagger}_\lambda(l)\right|.
\end{equation*}
To formulate the boundary condition in final form, we assume that the difference between phases of $\psi^{(-)}_\lambda$ and $\psi^{(+)}_\lambda$ near the interface is a multiple of $2\pi$:
\begin{equation*}
\varphi^{(+)}=\varphi^{(-)}+2\pi n, \hspace{0.25cm}n\in\mathbb{Z}.
\end{equation*}
As $l$ goes to zero, we obtain the following wavefunction-matching condition \cite{Kolesnikov2, Silin1}
\begin{equation*}
\sqrt{v^{(-)}_F}\psi^{(-)}_\lambda=\sqrt{v^{(+)}_F}\psi^{(+)}_\lambda,
\end{equation*}
where minus and plus signs refer to the materials on the left- and right-hand sides of the boundary, respectively.

The solution to Eq. \eqref{10} is expressed as follows.

{\bf 1.} At $x$ $<$ 0,
\begin{equation}\label{11}
\psi_\lambda(x)=C\begin{pmatrix}1\\q_1\end{pmatrix}e^{k_1x},
\end{equation}
where
\begin{equation*}
q_1=-i\frac{v_{F1}(k_1-\lambda k_y)}{E_\lambda-V_1+\Delta_1},
\end{equation*}
\begin{equation*}
v_{F1}k_1=\sqrt{\Delta^2_1-(E_\lambda-V_1)^2+v^2_{F1}k^2_y}.
\end{equation*}

{\bf 2.}  At 0 $<$ $x$ $<$ $d$,
\begin{equation}\label{12}
\begin{split}
\psi_\lambda(x)&=C\begin{pmatrix}\kappa^*\\q_2\kappa^*\end{pmatrix}e^{ik_2x}\\&+C\begin{pmatrix}\kappa\\-q_2\kappa\end{pmatrix}e^{-ik_2x},
\end{split}
\end{equation}
where
\begin{equation*}
\kappa=\frac{1}{2}\sqrt{\frac{v_{F1}}{v_{F2}}}\left[1+i\left(\frac{\lambda k_y}{k_2} +\frac{v_{F1}(k_1-\lambda k_y)E_\lambda}{v_{F2}k_2(E_\lambda-V_1+\Delta_1)}\right)\right],
\end{equation*}
\begin{equation*}
q_2=\frac{v_{F2}(k_2+i\lambda k_y)}{E_\lambda},\hspace{0.5cm}E_\lambda=\pm v_{F2}\sqrt{k^2_2+k^2_y},
\end{equation*}
with plus and minus corresponding to electrons and
holes, respectively.

{\bf 3.} At $x$ $>$ $d$,
\begin{equation}\label{13}
\psi_\lambda(x)=C\begin{pmatrix}\zeta\\q_3\zeta
\end{pmatrix}e^{-k_3(x-d)},
\end{equation}
where
\begin{equation*}
\begin{split}
\zeta&=\sqrt{\frac{v_{F1}}{v_{F3}}}\\ &\times\left[\cos(k_2d)+\left(\frac{\lambda k_y}{k_2}+\frac{v_{F1}(k_1-\lambda k_y)E_\lambda}{v_{F2}k_2(E_\lambda-V_1+\Delta_1)}\right)\sin(k_2d)\right],
\end{split}
\end{equation*}
\begin{equation*}
q_3=i\frac{v_{F3}(k_3+\lambda k_y)}{E_\lambda-V_3+\Delta_3},
\end{equation*}
\begin{equation*}
v_{F3}k_3=\sqrt{\Delta^2_3-(E_\lambda-V_3)^2+v^2_{F2}k^2_y}.
\end{equation*}
The constant $C$ is found by using the normalization condition for wavefunctions \eqref{11}---\eqref{13},
\begin{equation*}
\int\limits_{-\infty}^\infty\Psi_\lambda^{\dagger}(x)\Psi_\lambda(x)dx=1.
\end{equation*}

The carrier energy spectrum is determined by the dispersion relation
\begin{equation}\label{14}
\tan(k_2d)=v_{F2}k_2f(\lambda k_y; k_1, k_3, E_\lambda),
\end{equation}
where
\begin{equation*}
\begin{split}
f(\lambda k_y; &k_1, k_3, E_\lambda)=\left[v_{F1}(k_1-\lambda k_y)(E_\lambda-V_3+\Delta_3)\right.\\ &\left.+v_{F3}(k_3+\lambda k_y)(E_\lambda-V_1+\Delta_1)\right]\\
&\times\left[E_\lambda(E_\lambda-V_1+\Delta_1)(E_\lambda-V_3+\Delta_3)\right.\\
&-v_{F2}v_{F3}\lambda k_y(k_3+\lambda k_y)(E_\lambda-V_1+\Delta_1)\\
&+v_{F1}v_{F2}k_y(k_1-\lambda k_y)(E_\lambda-V_3+\lambda\Delta_3)\\
&\left.-v_{F1}v_{F3}(k_1-\lambda k_y)(k_3+\lambda k_y)E_\lambda\right]^{-1}
\end{split}
\end{equation*}
is a function of $k_2$ as well. Equation \eqref{14} must be solved for $k_2$, and then the energy  $E_\lambda$ is found.

In the case of an asymmetric quantum well, the dependence of \eqref{14} on $\lambda$ gives rise to pseudospin splitting as the extrema of the dispersion curves shift away from Brillouin-zone corners. The dispersion relation predicts that $E_\lambda(k_y)\neq E_{-\lambda}(k_y)$, and an energy splitting appears near the conduction-band bottom at $k_y=k^*_{ye}$:
\begin{equation*}
\delta E^e_s=|E^e_{-1}(k^*_{ye})-E^e_{+1}(k^*_{ye})|.
\end{equation*}

A similar energy splitting appears near the valence-band top at $k_y=k^*_{ye}$:
\begin{equation*}
\delta E^h_s=|E^h_{-1}(k^*_{yh})-E^h_{+1}(k^*_{yh})|
\end{equation*}
Thus, a graphene nanoribbon becomes an indirect band-gap semiconductor analogous to silicon and germanium, where an electron–hole plasma can exist \cite{Rice}. In the case of a symmetric quantum well ($\Delta_1=\Delta_3$, $V_1=V_3$, $v_{F1}=v_{F3}$) band structure is invariant under parity and there is no pseudospin splitting \cite{Ratnikov1}.

\begin{center}
4. SIZE QUANTIZATION
\end{center}

Solving Eq. \eqref{14}, we determine the size-quantized energies
\begin{equation*}
E_{\lambda b_\mp}(k_y)=\pm v_{F2}\sqrt{k^2_{2b_\mp}(\lambda k_y)+k^2_y},
\end{equation*}
where $b_\mp=1, 2, \ldots$ labels electron (--) and hole (+) branches, respectively. The size-quantized energy spectra for symmetric and asymmetric quantum wells are shown schematically in \hyperlink{Fig5}{Fig. 5}.

We now determine the carrier effective masses arising because of size quantization in the graphene nanoribbon in a planar heterostructure. Note that the effective masses are invariant under parity regardless of pseudospin splitting. Hereinafter, we omit indices $b_\mp$, restricting ourselves to a particular branch of the electron spectrum and a particular branch of the hole spectrum.

We write the dispersion law for electrons near an extremum at $\lambda k^*_{ye}$ as
\begin{equation}\label{15}
\begin{split}
E^e_\lambda&\approx E^e_0+\frac{1}{2m^*_e}\left(k_y-\lambda k^*_{ye}\right)^2,\\
m^*_e&=\frac{1}{v_{F2}}\frac{\sqrt{k^2_{20e}+k^{*2}_{ye}}}{1+k^{\prime2}_{20e}+k_{20e}k^{\prime\prime}_{20e}},
\end{split}
\end{equation}
where the respective values $k_{20e},\, k^\prime_{20e}$, $k^{\prime\prime}_{20e}$ of $k_{2e}(k_y)$ and its first and second derivatives at $k_y=\lambda k^*_{ye}$ are independent of $\lambda$; $E^e_0=v_{F2}\sqrt{k^2_{20e}+k^{*2}_{ye}}$ is the energy at the extremum.

\begin{figure}[!t]
\hypertarget{Fig5}{}
\begin{center}
\includegraphics[width=0.5\textwidth]{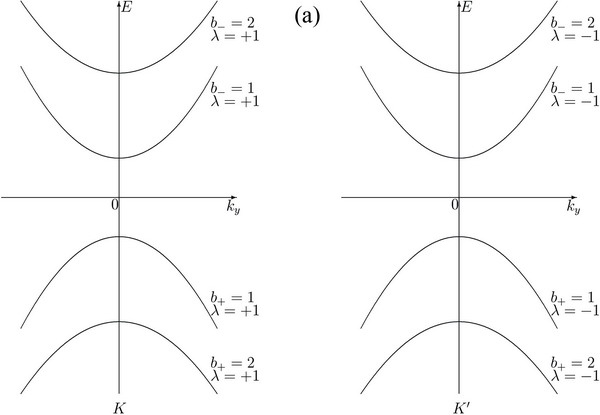}
\\
\vspace{0.25cm}
\includegraphics[width=0.5\textwidth]{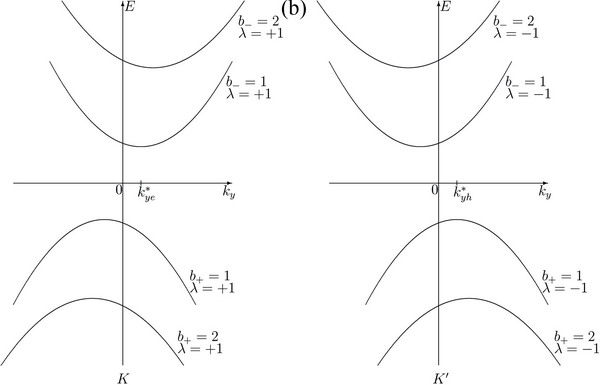}
\end{center}
{\bf Fig. 5.} Energy spectra: {\bf (a)} symmetric quantum well (no pseudospin splitting), with matching branches for $\lambda$ = +1 and $\lambda$ = --1 ($E_\lambda(k_y)$ = $E_{-\lambda}(k_y)$ = $E_\lambda(-k_y)$); {\bf (b)} asymmetric quantum well, with pseudospin splitting manifested by the “spread-out” in quasimomentum between the extrema at $k^*_{ye}$ for electrons, shown for $b_{-}$ = 1, and at $k^*_{yh}$ for holes, shown for $b_{+}$ = 1 ($E_\lambda(k_y)$ $\neq$ $E_{-\lambda}(k_y)$).
\end{figure}

Analogous expressions are obtained for hole energies:
\begin{equation}\label{16}
\begin{split}
E^h_\lambda&\approx E^h_0+\frac{1}{2m^*_h}\left(k_y-\lambda k^*_{yh}\right)^2,\\
m^*_h&=\frac{1}{v_{F2}}\frac{\sqrt{k^2_{20h}+k^{*2}_{ye}}}{1+k^{\prime2}_{20h}+k_{20e}k^{\prime\prime}_{20h}},
\end{split}
\end{equation}
where the respective values $k_{20h},\, k^\prime_{20h}$, $k^{\prime\prime}_{20h}$ of $k_{2h}(k_y)$ and its first and second derivatives at $k_y=-\lambda k^*_{yh}$; $E^h_0=-v_{F2}\sqrt{k^2_{20h}+k^{*2}_{yh}}$.

To estimate characteristic values, we consider the planar heterostructure combining a gapless nanoribbon with gapped graphene sheets with $\Delta_1=0.75$ eV, $v_{F1}=1.1v_{F2}$, $\Delta_3=1$ eV, and $v_{F3}=1.2v_{F2}$. The nanoribbon width is $d$ = 2.46 nm (ten hexagonal cells). Since the unknown values of $V_1$ and $V_3$ can be found by comparing our results with experimental data, we seek the dependence of energy spectrum parameters on $V_1$ and $V_3$. Note that $|V_1|\leq\Delta_1$ and $|V_3|\leq\Delta_3$ to ensure that the heterostructure is type I.

\begin{figure}[!t]
\hypertarget{Fig6}{}
\begin{center}
\includegraphics[width=0.5\textwidth]{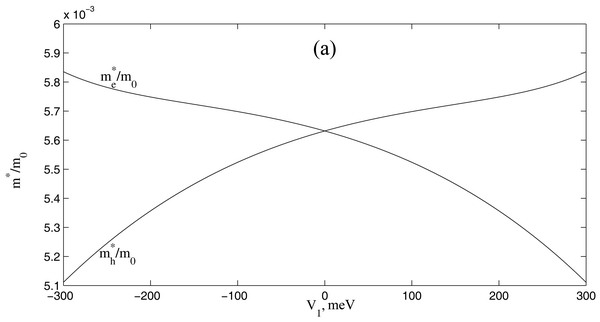}
\\
\vspace{0.25cm}
\includegraphics[width=0.5\textwidth]{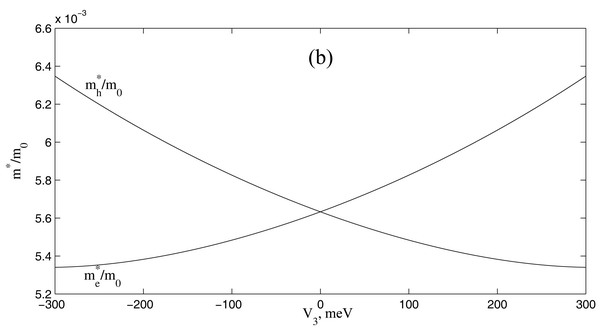}
\end{center}
{\bf Fig. 6.} Electron and hole effective masses in the graphene nanoribbon (in units of free-electron mass $m_0$) as functions of $V_1$ for $V_3$ = 0 {\bf(a)} and as functions of $V_3$ for $V_1$ = 0 {\bf(b)}.
\end{figure}

\hyperlink{Fig6}{Figures 6–9} show the results of numerical calculations of electron and hole effective masses in the graphe-ne nanoribbon, extremum energies, $k^*_{xe}$ and $k^*_{xh}$ values, and pseudospin splitting $\delta E^{e,h}_s$ plotted versus work function for one of the gapped graphene sheets given that the work function for the other is zero.

It is clear from Fig. 9 that the pseudospin splitting energy may amount to approximately 10 meV. To obtain a larger pseudospin splitting, the quantum well must be more asymmetric. Both $V_1$ and $V_3$ can be varied by shifting the valley energies in gapped graphene under applied stress, with potential barriers playing the role of bandgaps in the gapped graphene sheets. An analogous effect is achieved by applying an electric field on the order of $10^6$ V/cm perpendicular to the interfaces in the graphene plane \cite{Ratnikov2}.

As expected, the energy spectrum is symmetric under the change $E$ $\rightarrow$ $–E$ when $V_1$ and $V_3$ = 0; i.e., the electron and hole spectra have equal effective masses, extremum energies, extremum positions, and pseudo-spin splitting energies. The electron and hole effective masses in graphene are smaller than those in the gapped graphene sheets adjoining the gapless graphene nanoribbon ($m^*_1$ = $\Delta_1/v^2_{F1}$ $\approx$ $0.11m_0$ and $m^*_3$ = $\Delta_3/v^2_{F3}$ $\approx$ $0.15m_0$).

~

\vspace{0.25cm}
\begin{figure}[!t]
\begin{center}
\includegraphics[width=0.5\textwidth]{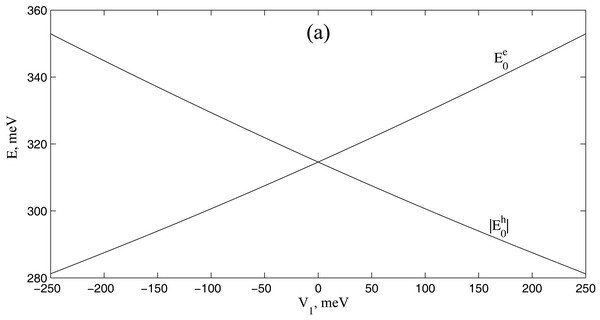}
\\
\vspace{0.25cm}
\includegraphics[width=0.5\textwidth]{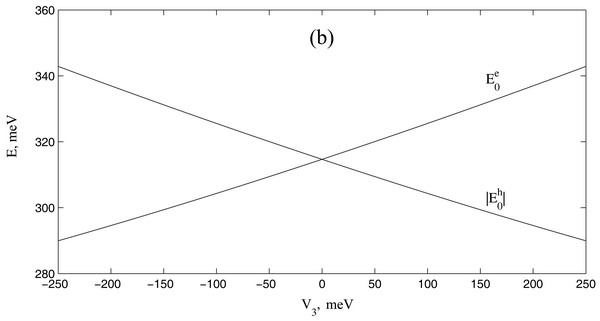}
\end{center}
\vspace{0.15cm}
{\bf Fig. 7.} Electron and hole extremum energies   and   in the size-quantization spectra as functions of $V_1$ for $V_3$ = 0 {\bf(a)} and as functions of $V_3$ for $V_1$ = 0 {\bf(b)}. The effective bandgap $E^{eff}_g$ = $E^e_0$ + $\left|E^h_0\right|$ $\approx$ 629 meV varies insignificantly.
\end{figure}

\begin{center}
5. INTERFACE STATES
\end{center}

The existence of surface states was predicted by Tamm in \cite{Tamm}. Extensive studies have been conducted of Tamm states in various systems including semiconductor superlattices \cite{Tikhodeev1, Tikhodeev2}. We consider interface states of a new type that arise in a narrow quasimomentum interval from the crossing of dispersion curves and are analogous to those in narrow-gap semiconductor heterostructures \cite{Kolesnikov3, Andyushin5}. In the planar graphene-based heterostructure examined here, these states are localized near the heterojunction interfaces between the nanoribbon and the gapped graphene sheets. Interface states can exist not only in quantum wells but also in quantum barriers \cite{Ando1}. Note that interface states arise as well from the crossing of dispersion curves in a single heterojunction between different graphene materials \cite{Ratnikov3}.

The wave function describing an interface electronic state is expressed as follows.

{\bf1.} At $x$ $<$ 0,

\begin{equation*}
\widetilde{\psi}_\lambda(x)=\widetilde{C}\begin{pmatrix}1\\ \widetilde{q}_1\end{pmatrix}e^{\kappa_1x},
\end{equation*}
where
\begin{equation*}
\widetilde{q}_1=-i\frac{u_1(\kappa_1-\lambda k_y)}{E_\lambda-V_1+\Delta_1},
\end{equation*}
\begin{equation*}
v_{F1}\kappa_1=\sqrt{\Delta^2_1-(E_\lambda-V_1)^2+v^2_{F1}k^2_y}.
\end{equation*}

\begin{figure}[!t]
\hypertarget{Fig8}{}
\begin{center}
\includegraphics[width=0.5\textwidth]{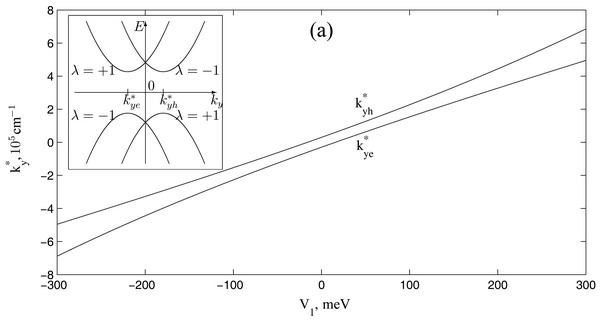}
\includegraphics[width=0.5\textwidth]{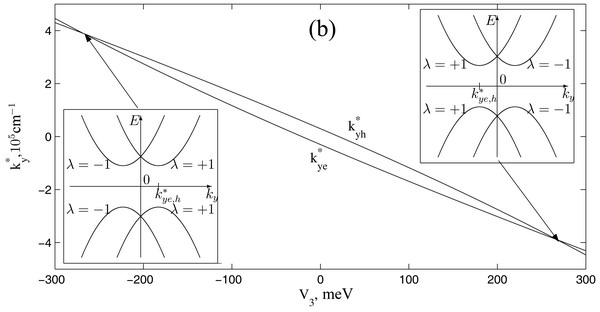}
\end{center}
\vspace{0.15cm}
{\bf Fig. 8.} Extremum points of size-quantization bran-ches for electrons ($k^*_y$ = $k^*_{ye}$) and holes ($k^*_y$ = $k^*_{yh}$) as functions $V_1$ for $V_3$ = 0 {\bf(a)} and as functions of $V_3$ for $V_1$ = 0 {\bf(b)}. Inserts show the relative positions of dispersion curves for $V_1$ = $V_3$ = 0 {\bf(a)} and at $k^*_{ye}$ = $k^*_{yh}$~{\bf(b)}; $K$ and $K^\prime$ points are set at the same  position for simplicity.
\end{figure}

~

{\bf2.} At 0 $<$ $x$ $<$ $d$,

\begin{equation*}
\widetilde{\psi}_\lambda(x)=\widetilde{C}\begin{pmatrix}\widetilde{\varkappa}_-\\
\widetilde{q}_2\widetilde{\varkappa}_-\end{pmatrix}e^{-\kappa_2x}+\widetilde{C}\begin{pmatrix}\widetilde{\varkappa}_+\\
\widetilde{q}^\prime_2\widetilde{\varkappa}_+\end{pmatrix}e^{\kappa_2x},
\end{equation*}
where
\begin{equation*}
\begin{split}
\widetilde{\varkappa}_\pm
&=\frac{1}{2}\sqrt{\frac{v_{F1}}{v_{F2}}}\left[1\pm\frac{\lambda
k_y}{\kappa_2}\pm \frac{v_{F1}(\kappa_1-\lambda
k_y)E_\lambda}{v_{F2}\kappa_2(E_\lambda-V_1+\Delta_1)}\right],\\
\widetilde{q}_2 &=i\frac{v_{F2}(\kappa_2+\lambda k_y)}{E_\lambda},\,
\widetilde{q}^\prime_2=-i\frac{v_{F2}(\kappa_2-\lambda
k_y)}{E_\lambda},
\end{split}
\end{equation*}
\begin{equation}\label{17}
E_\lambda=\pm v_{F2}\sqrt{k^2_y-\kappa^2_2},
\end{equation}
with plus and minus corresponding to electrons and holes,
respectively.

{\bf3.} At $x$ $>$ $d$

\begin{equation*}
\widetilde{\psi}_\lambda(x)=\widetilde{C}\begin{pmatrix}\widetilde{\zeta}\\
\widetilde{q}_3\widetilde{\zeta}\end{pmatrix}e^{-\kappa_3(x-d)},
\end{equation*}
where
\begin{equation*}
\begin{split}
&\widetilde{\zeta}=\sqrt{\frac{v_{F1}}{v_{F3}}}\\
&\times\left[ch(\kappa_2d)+\left(\frac{\lambda k_y}{\kappa_2}+
\frac{v_{F1}(\kappa_1-\lambda
k_y)E_\lambda}{v_{F2}\kappa_2(E_\lambda-V_1
+\Delta_1)}\right)sh(\kappa_2d)\right]
\end{split}
\end{equation*}
\begin{equation*}
\widetilde{q}_3=i\frac{v_{F3}(\kappa_3+\lambda
k_y)}{E_\lambda-V_3+\Delta_3},
\end{equation*}
\begin{equation*}
v_{F3}k_3=\sqrt{\Delta^2_3-(E_\lambda-V_3)^2+v^2_{F3}k^2_y}.
\end{equation*}

\begin{figure}[!t]
\begin{center}
\includegraphics[width=0.5\textwidth]{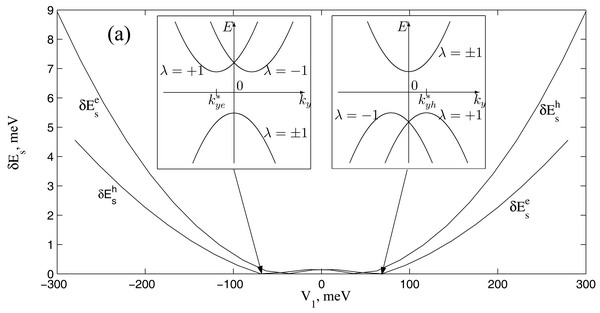}
\includegraphics[width=0.5\textwidth]{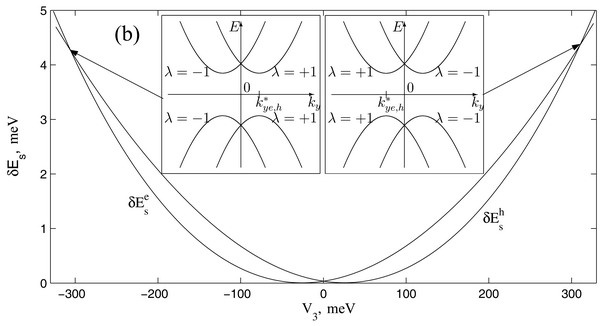}
\end{center}
\vspace{0.15cm} {\bf Fig. 9.} Pseudospin splitting in electron and
hole spectra, $\delta E^e_s$ and $\delta E^h_s$, as functions $V_1$
for $V_3$ = 0 {\bf(a)} and as functions of $V_3$ for $V_1$ = 0
{\bf(b)}. Vanishing $\delta E^{e,h}_s$ corresponds to vanishing
$k^*_{ye,h}$ in \hyperlink{Fig8}{Fig. 8}, as shown in inserts to
{\bf(a)}. Inserts to {\bf(b)} show positions of dispersion curves
when $k^*_{ye}$ and $k^*_{yh}$ coincide.
\end{figure}

The expression for energy in \eqref{17} implies that an interface
state exists only if\footnote{The zero mode corresponding to
$|\kappa_2|=|k_y|$ (with $E_\lambda$ = 0) is irrelevant here because
$\widetilde{\psi}_\lambda(x)$ $\equiv$ 0.}
\begin{equation*}
|\kappa_2|<|k_y|.
\end{equation*}

We obtain the dispersion relation
\begin{equation}\label{18}
\tanh(\kappa_2d)=v_{F2}\kappa_2f(\lambda k_y; \kappa_1, \kappa_3,
E_\lambda).
\end{equation}
which is similar to \eqref{14} up to the substitutions $k_1$
$\rightarrow$ $\kappa_1$, $k_2$ $\rightarrow$ $i\kappa_2$, and $k_3$
$\rightarrow$ $\kappa_3$.

We calculated numerically the energies of interface states as
functions of quasimomentum component $k_y$ for heterostructures with
$\Delta_1$ = 0.75 eV, $\Delta_3$ = 1 eV, $V_3$ = 0, $v_{F1}$ =
$v_{F3}$ = 0.7$v_{F2}$, and $V_1$ = 0 and 100 meV. The results shown
in \hyperlink{Fig10}{Figs. 10 and 11} demonstrate that only electron
interface states with $\lambda$ = +1 and hole interface states with
$\lambda$ = –1 occur.

When $V_1$ = 0, the allowed quasimomenta for hole interface states
($\lambda$ = –1) are similar to those for electron states, but the
hole and electron energies have opposite signs; i.e., the spectrum
is symmetric under the change $E_\lambda$ $\rightarrow$
$–E_\lambda$. When $V_1$ = 100 meV, the symmetry is broken and hole
interface states exist only at negative quasimomenta.

\begin{figure}[!t]
\hypertarget{Fig10}{}
\begin{center}
\includegraphics[width=0.5\textwidth]{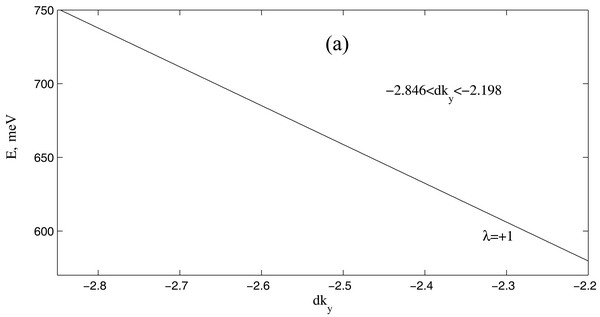}
\includegraphics[width=0.5\textwidth]{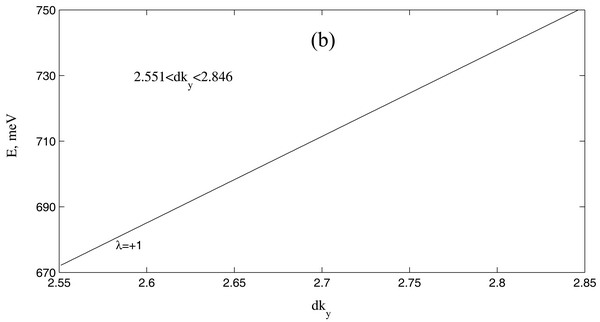}
\end{center}
\vspace{0.15cm} {\bf Fig. 10.} Energy of electron interface states
with $\lambda$ = +1 in the heterostructure with $\Delta_1$ = 0.75
eV, $\Delta_3$ = 1~eV, $v_{F1}$ = $v_{F3}$ = 0.7$v_{F2}$, and $V_1$
= $V_3$ = 0 (electron interface states with $\lambda$ = –1 do not
exist). Allowed quasimomenta lie in two intervals: {\bf(a)} negative
$k_y$ and {\bf(b)} positive $k_y$.
\end{figure}

\newpage

\begin{center}
6. EXCITON IN A PLANAR GRAPHENE\\ QUANTUM WELL
\end{center}

In gapless graphene, the carrier effective mass is zero and excitons
do not exist. The existence of excitons in gapless graphene would
lead to excitonic instability and excitonic insulator transition to
a gapped state \cite{Abrikosov, Brazovsky}.

The energy gap arising in a graphene nanoribbon as described in
Section 4 makes it possible to generate excitons by optical
excitation or electron-hole injection. Quantum-well excitons
strongly affect optical properties of the system considered here.

Excitons in similar quasi-one-dimensional carbon-based systems
(semiconducting single- and multi-walled nanotubes) have been
studied theoretically in \cite{Ando2}. The exciton spectrum is
calculated here for a planar graphene quantum well by using the
model applied to quantum wires in \cite{Babichenko}. This model
yields simple analytical expressions for exciton binding energy.

Since formulas \eqref{15} and \eqref{16} are obtained in the
nonrelativistic limit, the two-particle exciton wave function
depending on the electron and hole coordinates $y_-$ and $y_+$ in a
sufficiently narrow nanoribbon must obey the 1D Schr\"{o}dinger
equation with Coulomb potential:
\begin{equation}\label{19}
\begin{split}
&\left(-\frac{1}{2m^*_e}\frac{\partial^2}{\partial
y^2_-}-\frac{1}{2m^*_h}\frac{\partial^2}{\partial y^2_+}-
\frac{\widetilde{e}^2}{|y_--y_+|}\right)\phi(y_-, y_+)\\
&=E^\prime\phi(y_-, y_+),
\end{split}
\end{equation}
where $E^\prime=E-E^{eff}_g$ and $\widetilde{e}^2\equiv
e^2/\kappa_{eff}$. The effective dielectric constant of graphene,
$\kappa_{eff}=(\varepsilon+\varepsilon^\prime)/2$, may vary widely
with the dielectric constants $\varepsilon$ and $\varepsilon^\prime$
of the media in contact with graphene, such as free-space
permittivity and substrate dielectric constant \cite{Lozovik2,
Keldysh}.

\begin{figure}[!t]
\begin{center}
\includegraphics[width=0.5\textwidth]{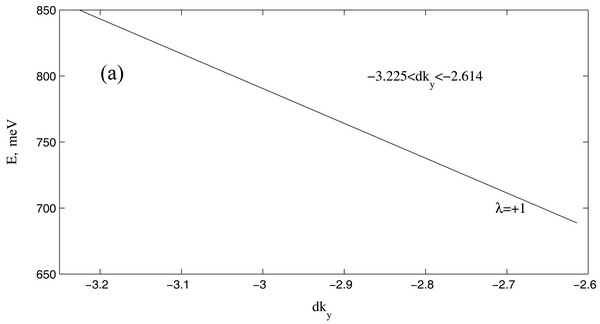}
\includegraphics[width=0.5\textwidth]{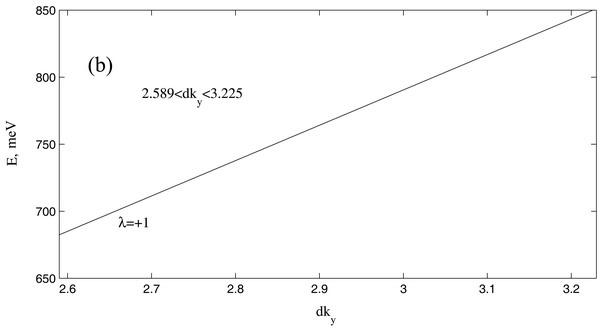}
\includegraphics[width=0.5\textwidth]{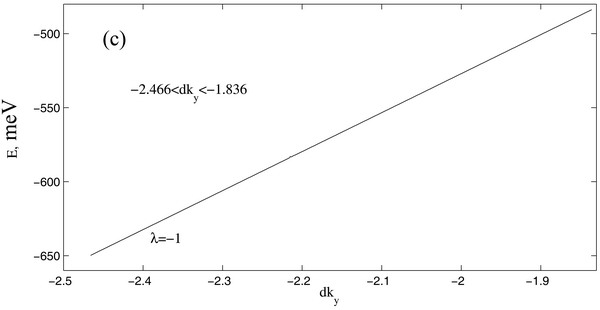}
\end{center}
\vspace{0.15cm} {\bf Fig. 11.} Same as in \hyperlink{Fig10}{Fig.
10}, but with $V_1$ = 100 meV; {\bf(c)} energy of hole states with
$\lambda$ = +1 for allowed quasimomenta in a single interval of
negative $k_y$.
\end{figure}

The electron--hole Coulomb interaction in a 1D gra-phene nanoribbon
is three-dimensional, but the problem can be reduced to one
dimension (electron and hole $y$ positions) for sufficiently narrow
nanoribbons.

Rewriting Eq. \eqref{19} in terms of electron–hole separation $y$ =
$y_--y_+$ and center-of-mass coordinate
\begin{equation*}
Y=\frac{m^*_ey_-+m^*_hy_+}{m^*_e+m^*_h}
\end{equation*}
and introducing the function
\begin{equation*}
\phi(y_-, y_+)=\psi_n(y)e^{iKY},
\end{equation*}
where $K$ is the total exciton momentum, we obtain
\begin{equation}\label{20}
\left(-\frac{1}{2\mu^*}\frac{\partial^2}{\partial
y^2}-\frac{\widetilde{e}^2}{|y|}\right)\psi_n(y)=E_n\psi_n(y),
\end{equation}
where $\mu^*=m^*_em^*_h/(m^*_e+m^*_h)$ is the reduced mass and $E_n$
is the energy of the $n$th exciton level ($n$ = 0, 1, 2, $\ldots$ is
the principal quantum number). The total exciton energy $E^\prime$
is obtained by adding the total kinetic energy of the electron–hole
pair to $E_n$:
\begin{equation*}
E^\prime=E_n+\frac{K^2}{2(m^*_e+m^*_h)}.
\end{equation*}

To find the solution at $y$ $>$ 0, we substitute $\psi_n(y)$
represented as
\begin{equation*}
\psi_n(y)=B_n\exp\left(-y/a_n\right)F_n\left(\frac{2y}{a_n}\right).
\end{equation*}
into Eq. \eqref{20} and obtain the confluent hypergeometric
differential equation
\begin{equation}\label{21}
\xi F^{\prime\prime}_n-\xi F^{\prime}_n+\eta F_n=0,
\end{equation}
where $\xi=\frac{2y}{a_n}$ and $\eta=\mu^*\widetilde{e}^2a_n$. We
also have
\begin{equation}\label{22}
E_n=-\frac{1}{2\mu^*a^2_n}.
\end{equation}
Equation \eqref{21} with $\eta$ = $n$ is solved by the associated
Laguerre polynomial
\begin{equation*}
F_n(\xi)=\frac{1}{n!}\xi
e^\xi\frac{d^n}{d\xi^n}\left(\xi^{n-1}e^{-\xi}\right)\equiv
L^{-1}_n(\xi).
\end{equation*}
and the wavefunction is expressed as
\begin{equation*}
\psi_n(y)=B_n\exp\left(-y/a_n\right)L^{-1}_n\left(\frac{2y}{a_n}\right).
\end{equation*}

Analogously, we find the solution to Eq. \eqref{20} at $y$~$<$~0:
\begin{equation*}
\psi_n(y)=\pm
B_n\exp\left(y/a_n\right)L^{-1}_n\left(-\frac{2y}{a_n}\right),
\end{equation*}
where ``+'' and ``–'' are taken for $n$ = 0 and $n$~$\neq$~0,
respectively, and the continuity of $\psi_n(y)$ and its first
derivative $\psi^\prime_n(y)$ are used as boundary conditions. Since
$\psi_n(0)$ = 0 and $\psi^\prime(0)$ $\neq$ 0 for $n$ $\neq$ 0, the
excited-state wavefunction $\psi_n(y)$ is odd (otherwise, it would
be discontinuous at the origin), whereas the ground-state
wavefunction is even.

The normalization condition
\begin{equation*}
\int\limits_{-\infty}^\infty|\psi_n(y)|^2dy=1
\end{equation*}
is used to determine the coefficient $B_n$ in the expression for
$\psi_n(y)$:
\begin{equation*}
B_n=\left[a_n\int\limits_0^\infty(L^{-1}_n(\xi))^2e^{-\xi}d\xi\right]^{-1/2}
\end{equation*}
and $B_n=1/\sqrt{2a_n}$ for $n=1,\, 2,\, \ldots$ and
$B_0=1/\sqrt{a_0}$ for $n$ = 0.

Here,
\begin{equation}\label{23}
a_n=\frac{n}{\mu^*\widetilde{e}^2}
\end{equation}
($n=1,\, 2\, \ldots$) is the Bohr radius of an exciton in the nth
excited state. Combining \eqref{22} with \eqref{23}, we find the
exciton energy spectrum:
\begin{equation}\label{24}
E_n=-\frac{\mu^*\widetilde{e}^4}{2n^2}.
\end{equation}

The 1D ground-state ($n$ = 0) Coulomb energy exhibits a logarithmic
divergence at short distances \cite{Loudon}. Therefore, the lateral
spread of the exciton wave function (along the $x$ axis) due to the
three-dimensional nature of Coulomb interaction should be taken into
account by introducing a cutoff parameter $d_0\lesssim d$. Averaging
the kinetic energy operator
\begin{equation*}
\widehat{T}=-\frac{1}{2\mu^*}\frac{\partial^2}{\partial y^2}
\end{equation*}
and the potential
\begin{equation*}
V(y)=-\frac{\widetilde{e}^2}{|y|}\theta\left(|y|-d_0\right)
\end{equation*}
over ground-state wave functions
\begin{equation}\label{25}
\psi_0(y)=\frac{1}{\sqrt{a_0}}e^{-|y|/a_0},
\end{equation}
where the ground-state Bohr radius $a_0$ plays the role of a
variational parameter, we express the ground-state exciton energy as
\begin{equation}\label{26}
E_0=\frac{1}{2\mu^*a^2_0}-\frac{2\widetilde{e}^2}{a_0}\ln\frac{a_0}{d}.
\end{equation}
Minimizing \eqref{26} with respect to $a_0$, we obtain an equation
for $a_0$:
\begin{equation}\label{27}
a_0=\frac{a_1}{2\left(\ln\frac{a_0}{d}-1\right)}.
\end{equation}
To logarithmic accuracy, when
\begin{equation}\label{28}
\ln\frac{a_1}{d}\gg1,
\end{equation}
we  find the relations
\begin{equation}\label{29}
E_0=4E_1\ln^2\frac{a_1}{d},
\end{equation}
\begin{equation}\label{30}
a_0=\frac{a_1}{2\ln\frac{a_1}{d}}.
\end{equation}
Using \eqref{27}, we easily obtain the next-order correction to
$E_0$:
\begin{equation*}
\delta
E^{(1)}_0=-8E_1\ln\frac{a_1}{d}\ln\left(2\ln\frac{a_1}{d}\right).
\end{equation*}
We now examine the applicability of the formulas derived here. The
semiconducting state induced in a graphene nanoribbon is stable with
respect to spontaneous electron--hole pair creation (excitonic
insulator transition) only if the exciton binding energy $|E_0|$ is
smaller than the effective bandgap in the graphene nanoribbon,
\begin{equation*}
|E_0|<E^{eff}_g.
\end{equation*}
Furthermore, the quantum well width $d$ must be much smaller than
the exciton Bohr radius $a_1$,
\begin{equation*}
d\ll a_1.
\end{equation*}
Logarithmically accurate formula \eqref{29} is correct only if
condition \eqref{28} holds. However, the asymmetric quantum well
analyzed here to examine pseudospin effects may not admit even a
single size-quantization level if the graphene nanoribbon width $d$
is too narrow. As $d$ decreases, the effective bandgap  increases,
approaching $\Delta_++\Delta_-$, where $\Delta_\pm$ =
$\min\{\Delta_1\pm V_1,\, \Delta_3\pm V_3\}$ (with plus and minus
corresponding to electrons and holes, respectively). When a certain
$d_c$ is reached, the size-quantization levels are pushed into the
continuum. This imposes a lower limit on $d$:
\begin{equation*}
d>d_c,
\end{equation*}
where $d_c$ can be estimated as
\begin{equation*}
d_c\simeq\frac{\pi v_{F2}}{\Delta_++\Delta_-}.
\end{equation*}
As $d$ increases, condition \eqref{28} is violated. In this case, a
more accurate variational calculation should be performed using the
modified three-dimensional Coulomb potential
\begin{equation*}
\widetilde{V}(y)=-\frac{\widetilde{e}^2}{\sqrt{y^2+d^2_0}},
\end{equation*}
where $d_0$ is a cutoff parameter. We average the Hamiltonian with
potential $\widetilde{V}(y)$ over trial functions \eqref{25} to
obtain
\begin{equation}\label{31}
E_0=\frac{1}{2\mu^*a^2_0}-\frac{2\widetilde{e}^2}{a_0}I(\rho),
\end{equation}
where $I(\rho)=\frac{\pi}{2}\left[H_0(\rho)-Y_0(\rho)\right]$,
$H_\nu(\rho)$ is a Struve function, $Y_\nu(\rho)$ is a Bessel
function of the second kind, and $\rho=2d_0/a_0$ ($\nu$ is 0 here
and 1 below).

Minimizing \eqref{31} with respect to $a_0$, we obtain an equation
for $a_0$:
\begin{equation}\label{32}
\frac{2a_0}{a_1}I(\rho)+ \frac{4d_0}{a_1}J(\rho)=1,
\end{equation}
where $J(\rho)=1-\frac{\pi}{2}\left[H_1(\rho)-Y_1(\rho)\right]$.

\hyperlink{Fig12}{Figure 12} shows the numerical results obtained by
using both methods to calculate $E_0(d)$ for the heterostructure
described in Section 4, with $d_0$ $\propto$ $d$ adjusted to match
the curves at small $d$. Discrepancy at large $d$ increases as
$\ln(a_1/d)$ approaches unity.

\begin{center}
7. ELECTRIC FIELD EFFECT\\ ON EXCITON LEVELS
\end{center}

Interaction between an exciton and an external electrostatic field
$\boldsymbol{\mathcal{E}}$ is described by the operator
\begin{equation*}
\widehat{H}_i=-{\bf
d}\boldsymbol{\mathcal{E}}=|\widetilde{e}|(\mathcal{E}_xx+\mathcal{E}_yy)
\end{equation*}
where $x$ = $|x_-- x_+|$ and $y$ = $|y_-- y_+|$ are the
electron–hole relative position vector components and ${\bf d}$ is
dipole moment. The electric field is supposed to be weak enough to
ensure that the energy level shift is not only smaller than the
spacing between size-quantization levels but also smaller than the
spacing between exciton levels. These conditions can be written as
\begin{equation*}
d\ll a_1\ll a_\mathcal{E},
\end{equation*}
where $a_\mathcal{E}=(\mu^*|\widetilde{e}|\mathcal{E})^{-1/3}$ is
the electric length.

\begin{figure}[!t]
\hypertarget{Fig12}{}
\begin{center}
\includegraphics[width=0.5\textwidth]{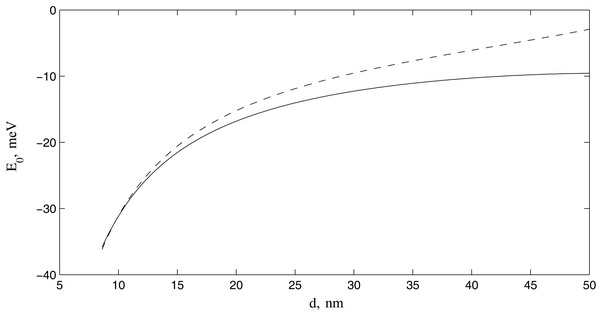}
\end{center}
\vspace{0.15cm} {\bf Fig. 12.} Exciton ground-state energy
calculated by formula \eqref{29} (dashed line) and by formula
\eqref{31} after Eq. \eqref{32} is solved numerically for $a_0$
(solid line); $d_0$ = 0.22$d$.
\end{figure}

We consider two cases: (1) the electric field is applied parallel to
the $x$ axis and perpendicular to the nanoribbon edges in the
graphene plane; (2) the electric field is applied along the $y$
axis, parallel to the nanoribbon edges.

In the former case, the energy shift varies linearly with the
difference between the average $x$ components of the electron and
hole position vectors:
\begin{equation}\label{33}
E^{(1)}_{\bot\lambda\lambda^\prime}=|\widetilde{e}|\mathcal{E}\left(\langle
x_-\rangle_\lambda-\langle x_+\rangle_{\lambda^\prime}\right),
\end{equation}
where average x components are calculated by using electron and hole
single-particle wave functions, generally depending on the electron
and hole eigenvalues $\lambda$ and $\lambda^\prime$ of the operator
$\widehat{P}$, respectively. Exciton energy shift \eqref{33} is
independent of the principal quantum number $n$. It may vary with
$\lambda$ and $\lambda^\prime$, resulting in different exciton
binding energies (more precisely, the binding energy of an
electron--hole pair with $\lambda$ = $\pm1$ and
$\lambda^\prime$~=~$\pm1$ may have four different values).

In the latter case, the first-order electric field-induced
correction is zero\footnote{Note that $E^{(1)}_{\bot n}\equiv0$ in
the former case if the electron and hole spectra transform into each
other under field inversion.},
\begin{equation}\label{34}
E^{(1)}_{\|n}=|\widetilde{e}|\mathcal{E}\langle y\rangle_n\equiv0,
\end{equation}
because the integral of $y|\psi_n(y)|^2$ with respect to $y$
vanishes. To evaluate the second-order electric field-induced
correction, we make use of the Dalgarno--Lewis perturbation theory
\cite{Dalgarno}. Defining a Hermitian operator such that
\begin{equation}\label{35}
[\widehat{F},\ \widehat{H}_0]|n\rangle=\widehat{H}_i|n\rangle,
\end{equation}
where
\begin{equation*}
\widehat{H}_0=-\frac{1}{2\mu^*}\frac{\partial^2}{\partial
y^2}-\frac{\widetilde{e}^2}{|y|}
\end{equation*}
is the zeroth-order Hamiltonian, $|n\rangle=\psi_n(y)$ is the
zeroth-order wave function of the $n$th exciton level, and
$\widehat{H}_i=|\widetilde{e}|\mathcal{E}y$, we obtain
\begin{equation}\label{36}
E^{(2)}_{\|n}=\langle n|\widehat{H}_i\widehat{F}|n\rangle-\langle
n|\widehat{H}_i|n\rangle\langle n|\widehat{F}|n\rangle.
\end{equation}
In the case in question, the second term in this formula vanishes by
virtue of \eqref{34}.

Rewriting Eq. \eqref{35} as
\begin{equation}\label{37}
\psi_n\frac{\partial^2\widehat{F}}{\partial
y^2}+2\frac{\partial\psi_n}{\partial
y}\frac{\partial\widehat{F}}{\partial y}=2\mu^*\widehat{H}_i\psi_n,
\end{equation}
we find
\begin{equation}\label{38}
\begin{split}
&\widehat{F}(y)=2\mu^*\int\limits_{-\infty}^y\frac{dy^\prime}{|\psi_n(y^\prime)|^2}\\
&\times\int\limits_{-\infty}^{y^\prime}dy^{\prime\prime}\psi^*_n(y^{\prime\prime})\widehat{H}_i\psi_n(y^{\prime\prime}).
\end{split}
\end{equation}

Combining \eqref{36} with \eqref{38}, we have the exciton
gro-und-state energy shift
\begin{equation}\label{39}
E^{(2)}_{\|0}=-\frac{5}{128}\frac{a^3_1}{\ln^4\frac{a_1}{d}}\mathcal{E}^2,
\end{equation}
which is very small compared to $E_0$ given by \eqref{29} because of
the fourth power of a logarithm in the denominator and a small
numerical factor.

For comparison, we write out the energy correction to the first
excited exciton state:
\begin{equation*}
E^{(2)}_{\|1}=-\frac{3}{8}(31-6\gamma)a^3_1\mathcal{E}^2,
\end{equation*}
where $\gamma$ = 0.577\ldots\, is Euler’s constant.

By analogy with layered heterostructures \cite{Miller}, the ionizing
(exciton-breaking) field strength $\mathcal{E}_c$ is estimated as
\begin{equation}\label{40}
\mathcal{E}_c=\frac{|E_0|}{8|\widetilde{e}|\langle|y|\rangle_0},
\end{equation}
where $\langle|y|\rangle_0=a_0/2$ is the average electron–hole
separation for the ground-state exciton. To logarithmic accuracy, it
follows that
\begin{equation}\label{41}
\mathcal{E}_c=\mu^{*2}|\widetilde{e}|^5\ln^3\frac{a_1}{d}.
\end{equation}

To get the order of magnitude of $\mathcal{E}_c$, consider the
quantum well discussed in Section 4. Setting $m^*_e$ = $m^*_h$
$\approx$ 0.0056$m_0$, the SiO$_2$ substrate dielectric constant
$\kappa_{eff}$ $\approx$ 5, $d$ = 2.46 nm, and $a_1$ $\approx$ 81
nm, we use formula~\eqref{41} to obtain $\mathcal{E}_c$ = 9 kV/cm.

\begin{center}
8. POSSIBLE EXPERIMENTS\\ ON THE HETEROSTRUCTURE
\end{center}

Pseudospin splitting can be observed by means of Raman spectroscopy.
The $D^\prime$ peak of interest for the present study (alternatively
called $2D$ peak to emphasize that it is due to a
two-phonon-assisted process) is located at 2700 cm$^{–1}$
\cite{Graf}. It arises from intervalley scattering involving phonons
with wavenumbers $q$~$>$~$K$, where $K$ = $4\pi/3\sqrt{3}a$
$\approx$ $1.7\times10^8$ cm$^{-1}$ is the spacing between adjacent
$K$ and $K^\prime$ points. One process of this kind is indicated as
$A$ $\rightarrow$ $B$ $\rightarrow$ $C$ $\rightarrow$ $D$
$\rightarrow$ $A$ in \hyperlink{Fig13}{Fig. 13}.

\begin{figure}[!t]
\hypertarget{Fig13}{}
\begin{center}
\includegraphics[width=0.5\textwidth]{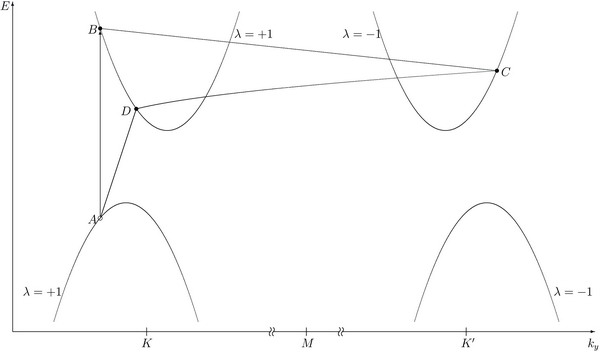}
\end{center}
\vspace{0.15cm} {\bf Fig. 13.} Possible double resonant Raman
processes involving electron scattering between valleys. To simplify
presentation, analogous processes involving hole scattering between
valleys are not shown.
\end{figure}

Pseudospin splitting enables intervalley scattering involving
phonons with $q^\prime$ $\approx$ $q\mp\Delta k$ (with plus for
electrons and minus for holes), where $\Delta k$ = $2k^*_{ye}$ and
$\Delta k$~=~$2k^*_{yh}$ in electron and hole scattering,
respectively. These processes contribute to a peak blueshifted from
$D^\prime$ by $\Delta\omega^{(+)}_R$ and a peak redshifted from
$D^\prime$ by $\Delta\omega^{(-)}_R$, giving rise to a doublet
structure of the $D^\prime$ peak.

An estimate for $\Delta\omega_R$ can be obtained by using optical
phonon dispersion $\omega_{ph}(q)$. The Raman shift is twice the
optical phonon frequency:
\begin{equation*}
\delta\omega_R(q)=2\omega_{ph}(q).
\end{equation*}
The change in the Raman shift caused by pseudospin splitting is
\begin{equation*}
\Delta\omega^{(\pm)}_R\approx|\delta\omega_R(\Delta K\mp\Delta
k)-\delta\omega_R(\Delta K)|.
\end{equation*}
which amounts to $\Delta\omega_R\approx24$ cm$^{-1}$ for
characteristic values of the heterostructure parameters. This value
essentially exceeds the Raman spectral resolution of 1 cm$^{–1}$
\cite{Ni} and compares to the $D^\prime$ peak width for gapless
graphene, $\Gamma_0=30$ cm$^{-1}$ \cite{Ferrari, Calizo}.

Note that a blue shift of the $D^\prime$ peak has also been observed
in the Raman spectrum of epitaxial graphene on a SiC substrate
\cite{Ni}. This effect is attributed to the strain induced by the
substrate in quasi-free graphene since the SiC lattice constant
exceeds substantially that of graphene.

Raman scattering contributions from gapped gra-phene sheets can be
avoided either by using a laser beam whose width is smaller than
that of the gapless graphene nanoribbon ($d$ $\lesssim$ 10 nm) or by
pumping at a frequency $\omega$ such that the beam cannot be
absorbed by gapped graphene materials,
\begin{equation*}
E^{eff}_g+2\omega_{ph}<\omega<min\{2\Delta_1,\, 2\Delta_3\}
\end{equation*}

The positions of the luminescence lines corresponding to exciton
levels can be determined from optical experiments and compared to
theoretical predictions. The splitting of exciton lines in an
electric field is evaluated by using formulas \eqref{33} and
\eqref{39}, respectively.

Let us now discuss the Landau levels in a planar graphene
heterostructure induced by a magnetic field perpendicular to the
graphene plane. The magnetic field is supposed to be relatively
weak,
\begin{equation*}
a_H\,=\,\sqrt{c/|e|H}\gg d,
\end{equation*}
so that size quantization plays a dominant role.

As in analyses of layered narrow-gap heterostructures
\cite{Andryushin3, Silin2}, the quasimomenta of free-moving carriers
in size-quantized spectra \eqref{15} and \eqref{16} are replaced by
the quantized magnetic momentum
\begin{equation*}
k_{Hn}=\frac{\sqrt{2n}}{a_H},
\end{equation*}
where $n=0,\, 1,\, 2\,\ldots$ labels Landau levels. (While each
size-quantization level splits into a respective set of Landau
levels, the present analysis is restricted to the lowest
size-quantization level.)

Then, we have the carrier energies
\begin{equation*}
\begin{split}
E^e_{n\lambda}&=E^e_0+\omega^e_c\left(\sqrt{n}-\lambda\sqrt{n_{He}}\right)^2,\\
E^h_{n\lambda}&=E^h_0-\omega^h_c\left(\sqrt{n}+\lambda\sqrt{n_{Hh}}\right)^2,
\end{split}
\end{equation*}
where $\omega^{e,h}_c=\frac{|e|H}{m^*_{e,h}c}$ is the cyclotron
frequency and $n_{He,h}=a^2_Hk^{*2}_{ye,h}/2$.

The minimum energy of electrons (near $K$ point, with $\lambda$ =
+1) or holes (near $K^\prime$ point, with $\lambda$ = –1) may match
a Landau level with $n$ $\neq$ 0. The zeroth Landau level is lowest
only if $n_{He,h}$ $<$ 1/4. When
\begin{equation*}
N_{n-1}<n_{He,h}<N_n,
\end{equation*}
where $N_n=\left(\sqrt{n+1}+\sqrt{n}\right)^2/4$, the minimum energy
(ground state) matches the nth Landau level.

Such an unusual ground state can be observed only at temperatures
sufficiently low that the excitation energy from the lowest Landau
level to the next one satisfies the condition $\Delta E^{e,h}_H> T$.
It can easily be shown that $\Delta E^{e,h}_H$ as a function of
$n_{He,h}$ has minima at $n_{Hmin}=N_n$ ($n=0,\, 1,\, 2\,\ldots$)
and maxima at $n_{Hmax}=\left(\sqrt{n+1}+\sqrt{n-1}\right)^2/4$
($n=1,\, 2\,\ldots$):
\begin{equation*}
\begin{split}
&\Delta
E^{e,h}_H\\&=\begin{cases}\omega^{e,h}_c\left[\left(\sqrt{n-1}-\sqrt{n_{He,h}}\right)^2-
\left(\sqrt{n}-\sqrt{n_{He,h}}\right)^2\right],\\
N_{n-1}<n_{He,h}<n_{Hmax},\\
\omega^{e,h}_c\left[\left(\sqrt{n+1}-\sqrt{n_{He,h}}\right)^2-
\left(\sqrt{n}-\sqrt{n_{He,h}}\right)^2\right],\\
n_{Hmax}<n_{He,h}<N_n.
\end{cases}
\end{split}
\end{equation*}
A degenerate ground state corresponds to zero excitation energy. The
excitation energy $\Delta E^{e,h}_H$ is plotted as a function of
$n_{He,h}$ in \hyperlink{Fig14}{Fig. 14}.

Finally, interface states can manifest themselves in the I–V curve
of the planar heterostructure carrying a current parallel to the
gapless graphene nanoribbon. An increase in applied electric field
may cause charge carriers to ``drop'' into interface states
(preferable ener-gy-wise), giving rise to a region of negative
differential conductivity in the I–V curve.

\begin{figure}[!t]
\hypertarget{Fig14}{}
\begin{center}
\includegraphics[width=0.5\textwidth]{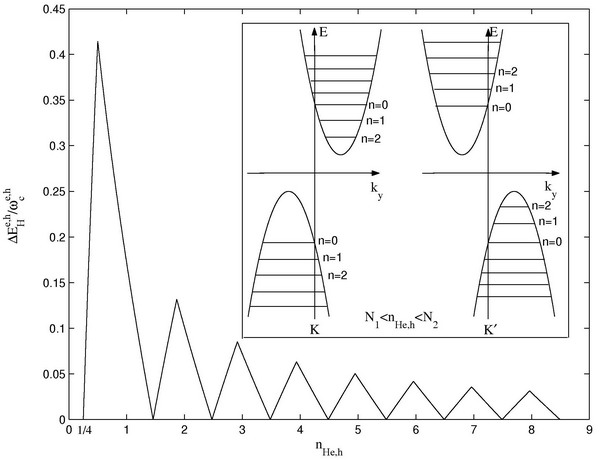}
\end{center}
\vspace{0.15cm} {\bf Fig. 14.} Excitation energy $\Delta E^{e,h}_H$
vs. $n_{He,h}$. The insert shows positions of Landau levels with
$N_1$~$<$~$n_{He,h}$~$<$~$N_2$.
\end{figure}

\begin{center}
8. CONCLUSIONS
\end{center}

We have analyzed the characteristics of planar gra-phene
nanostructures (quantum wells). On the one hand, they retain the
unique properties of infinite gra-phene sheets. On the other hand,
bandgap opening makes them important building blocks in carbon-based
nanoelectronics, which can be used to control electron motion.

Parameters of graphene quantum wells can easily be manipulated by
varying the gapless nanoribbon width or the potential barriers in
the adjoining gapped gra-phene sheets. We predict pseudospin
splitting to occur in asymmetric graphene quantum wells and
interface states to arise from the crossing of dispersion curves of
gapless and gapped graphene materials.

We have performed calculations of optical properties of planar
graphene nanostructures and suggested possible experiments to study
the effects in question.

Analysis of pseudospin (valley) characteristics in the
heterostructure is simplified by using an effective Hamiltonian
having a pseudospin-split energy spectrum. Note that an analogous
spectrum was discussed in \cite{Rashba1, Rashba2}. Therefore, the
effective Hamiltonian must contain a Bychkov–Rashba spin-orbit
coupling $\widehat{H}_R=\alpha_R[{\boldsymbol\sigma}\widehat{\bf
p}]{\boldsymbol\nu}$ responsible for pseudospin splitting, where
$\alpha_R$ is a Rash-ba parameter and ${\boldsymbol\nu}$ is the unit
vector normal to the heterojunction interface in the graphene plane.
In the coordinate system used in this paper, effective electron and
hole Hamiltonians can be written as
\begin{equation*}
\widehat{H}^e_{eff}=\frac{\widehat{p}^2_y}{2m^*_e}-\lambda\alpha_{Re}\sigma_z\widehat{p}_y+\Delta_e
\end{equation*}
and
\begin{equation*}
\widehat{H}^h_{eff}=\frac{\widehat{p}^2_y}{2m^*_h}+\lambda\alpha_{Rh}\sigma_z\widehat{p}_y+\Delta_h,
\end{equation*}
where $\alpha_{Re}=k^*_{ye}/m^*_e$ and $\alpha_{Rh}=k^*_{yh}/m^*_h$
re the respective electron and hole Rashba parameters (which are
different when the symmetry under field inversion is broken) and the
parameters $\Delta_{e,h}$ are expressed in terms of extremum
energies and Rashba parameters as
\begin{equation*}
\begin{split}
\Delta_e&=E^e_0+\frac{1}{2}m^*_e\alpha^2_{Re},\\
\Delta_h&=-E^h_0+\frac{1}{2}m^*_h\alpha^2_{Rh}.
\end{split}
\end{equation*}
For characteristic values of the heterostructure parameters, our
estimates predict $\alpha_{Re,h}\simeq5\cdot10^{-8}$ eV$\cdot$cm,
which is higher by two orders of magnitude than for GaAs
\cite{Rashba1} (mainly because of small effective mass).

\begin{center}
ACKNOWLEDGMENTS
\end{center}

This work was supported by the Dynasty Foundation, the Scientific
and Educational Complex of the Lebedev Physical Institute, and by
the Presidium of the Russian Academy of Sciences under the Program
for the Support of Young Scientists.


\begin{thebibliography}{70}
\bibitem{Novoselov1}
K.S. Novoselov, A.K. Geim, S.V. Morozov et al., Science {\bf 306}, 666 (2004).
\bibitem{Novoselov2}
K.S. Novoselov, A.K. Geim, S.V. Morozov et al., Nature {\bf 438}, 197 (2005).
\bibitem{Zhang1}
Y. Zhang, Y.-W. Tan, H. L. Stormer, and P. Kim, Nature {\bf 438}, 201 (2005).
\bibitem{Morozov}
S.V. Morozov, K.S. Novoselov, M.I. Katsnelson et al., Phys. Rev. Lett. {\bf 100}, 016602 (2008).
\bibitem{Du}
X. Du, I. Skachko, A. Barker, and E.Y. Andrei, Nat. Nanotech. {\bf 3}, 491 (2008).
\bibitem{Brey1}
L. Brey and H.A. Fertig, Phys. Rev. B {\bf 73}, 235411 (2006).
\bibitem{Brey2}
L. Brey and H.A. Fertig, Phys. Rev. B {\bf 75}, 125434 (2007).
\bibitem{Son}
Y.-W. Son, M.L. Cohen, and S.G. Louie, Phys. Rev. Lett. {\bf 97},
216803 (2006).
\bibitem{Saito}
R. Saito, G. Dresselhaus, and M.S. Dresselhaus, \textsl{Physical Properties of Carbon Nanotubes}, (Imperial College Press, London, 1998).
\bibitem{Ando1}
T. Ando, J. Phys. Soc. Jpn. {\bf 74}, 777 (2005).
\bibitem{Wang}
X. Wang, Y. Ouyang, X. Li et al., Phys. Rev. Lett. {\bf 100}, 206803 (2008).
\bibitem{Ponomarenko}
L.A. Ponomarenko, F. Schedin, M.I. Katsnelson et al., Science {\bf 320}, 356 (2008).
\bibitem{Giovannetti}
G. Giovannetti, P.A. Khomyakov, G. Brocks et al., Phys. Rev. B {\bf 76}, 073103 (2007).
\bibitem{Mattausch}
A. Mattausch and O. Pankratov, Phys. Rev. Lett. {\bf 99}, 076802 (2007).
\bibitem{Zhou}
S.Y. Zhou, G.-H. Gweon, A.V. Fedorov et al., Nature Mater. {\bf 6}, 770 (2007).
\bibitem{Elias}
D.C. Elias, R.R. Nair, T.M.G. Mohiuddin et al., Science {\bf 323}, 610 (2009).
\bibitem{Lebegue}
S. Leb\`{e}gue, M. Klintenberg, O. Eriksson, and M.I. Katsnelson, Phys. Rev. B {\bf 79}, 245117 (2009).
\bibitem{Zanella}
I. Zanella, S. Guerini, S.B. Fagan et al., Phys. Rev. B {\bf 77}, 073404 (2008).
\bibitem{Marchini}
S. Marchini, S. G$\ddot{u}$nther, and J. Witterlin, Phys. Rev. B {\bf 76}, 075429 (2007).
\bibitem{Martoccia}
D. Martoccia, P.R. Willmon, T. Brugger et al., Phys. Rev. Lett. {\bf 101}, 126102 (2008).
\bibitem{Pletikosic}
I. Pletikosi\'{c}, M. Kralj, P. Pervan et al., Phys. Rev. Lett. {\bf 102}, 056808 (2009).
\bibitem{Han}
M.Y. Han, B. \"{O}zyilmaz, Y. Zhang, and P. Kim, Phys. Rev. Lett. {\bf 98}, 206805 (2007).
\bibitem{Vorobiev}
L.E. Vorob’ev, E.L. Ivchenko, D.A. Firsov, and V.A . Shalygin ,  \textsl{Optical Properties of Nanostructures} (Nauka, St. Petersburg, 2001) [in Russian].
\bibitem{Ratnikov1}
P.V. Ratnikov and A.P. Silin, Kratk. Soobshch. Fiz., No. 2, 11 (2009); \href{http://arxiv.org/pdf/0808.3388}{arXiv:0808.3388}.
\bibitem{Idlis}
B.G. Idlis and M.Sh. Usmanov, Sov. Phys. Semicond. {\bf 26} (2), 186 (1992).
\bibitem{Akchiezer}
A. I. Akhiezer and V. B. Berestetskii, \textsl{Quantum Electrodynamics}  (Wiley, New York, 1965; Nauka, Moscow, 1969).
\bibitem{Neto1}
A.H. Castro Neto, F. Guinea, N.M.R. Peres et al., Rev. Mod. Phys. {\bf 81}, 109 (2009).
\bibitem{Lozovik1}
Yu.E. Lozovik, S.P. Merkulova, and A.A. Sokolik, Phys.--Usp. {\bf 51} (7), 727 (2008).
\bibitem{Schweber}
S.S. Schweber, \textsl{Introduction to Relativistic Quantum Field Theory} (Halper and Row, New York, 1961).
\bibitem{Tsvelik}
A.M. Tsvelik, \textsl{Quantum Field Theory in Condensed Matter Physics} (Cambridge University Press, 1998).
\bibitem{Appelquist}
T.W. Appelquist, M. Bowick, D. Karabali, and L.C.R. Wijewardhana, Phys. Rev. D {\bf 33}, 3704 (1986).
\bibitem{Volkov}
B.A. Volkov, B.G. Idlis, and M.Sh. Usmanov, Phys.--Usp. 38 (7), 761 (1995).
\bibitem{Kolesnikov1}
A.V. Kolesnikov and A.P. Silin, JETP {\bf 82} (6), 1145 (1996).
\bibitem{Kolesnikov2}
A.V. Kolesnikov and A.P. Silin, J. Phys.: Condens. Matter {\bf 9}, 10929 (1997).
\bibitem{Silin1}
A.P. Silin and S.V. Shubenkov, Phys. Solid State {\bf 40} (7), 1223 (1998).
\bibitem{Andryushin1}
E.A. Andryushin, Sh.U. Nutsalov, and A.P. Silin, Phys. Low-Dim. Struct. {\bf 7/8}, 85 (1999).
\bibitem{Andryushin2}
E.A. Andryushin, S.A. Vereshchagin, and A.P. Silin, Kratk. Soobshch. Fiz., No. 6, 21 (1999).
\bibitem{Andryushin3}
E.A. Andryushin, A.P. Silin, and S.A. Vereshchagin, Phys. Low-Dim. Struct. {\bf 3/4}, 85 (2000).
\bibitem{Andryushin4}
E.A. Andryushin, Sh.U. Nutsalov, and A.P. Silin, Kratk. Soobshch. Fiz., No. 3, 3 (2001).
\bibitem{Ratnikov2}
P.V. Ratnikov and A.P. Silin, Kratk. Soobshch. Fiz., No. 11, 22 (2005).
\bibitem{Rycerz}
A. Rycerz, J. Tworzyd{\l}o, and C.W.J. Beenakker, Nature Phys. {\bf 3}, 172 (2007).
\bibitem{Tworzydlo}
J. Tworzyd{\l}o, I. Snyman, A.R. Akhmerov, and C.W.J. Beenakker, Phys. Rev. B {\bf 76}, 035411 (2007).
\bibitem{Akhmerov}
A.R. Akhmerov and C.W.J. Beenakker, Phys. Rev. Lett. {\bf 98}, 157003 (2007).
\bibitem{Xiao}
D. Xiao, W. Yao, and Q. Niu, Phys. Rev. Lett. {\bf 93}, 236809 (2007).
\bibitem{Zhang2}
Z.Z. Zhang, K. Chang, and K.S. Chan, Applied Phys. Lett. {\bf 93}, 062106 (2008).
\bibitem{Carcia}
J.L. Carcia-Pomar, A. Cortijo, and M. Nieto-Vesperinas, Phys. Rev. Lett. {\bf 100}, 236801 (2008).
\bibitem{Pereira}
J.M. Pereira Jr., F.M. Peeters, R.N. Costa Filho, and G.A. Farias, J. Phys.: Condens. Matter {\bf 21}, 045301 (2009).
\bibitem{Semenoff}
G.W. Semenoff, Phys. Rev. Lett. {\bf 53}, 2449 (1984).
\bibitem{Abanin}
D.A. Abanin, P.A. Lee, and L.S. Levitov, Phys. Rev. Lett. {\bf 96}, 176803 (2006).
\bibitem{Rice}
T.M. Rice, J.C. Hensel, T. . Phillips, and G.A. Thomas, \textsl{The Electron–Hole Liquid in Semiconductors: Theoretical Aspects} (Academic, New York, 1977; Mir, Moscow, 1980).
\bibitem{Suzuura}
H. Suzuura and T. Ando, Phys. Rev. Lett. {\bf 89}, 266603 (2002).
\bibitem{Tamm}
I.E. Tamm, Phys. Z. Sowjetunion {\bf 1}, 733 (1932).
\bibitem{Tikhodeev1}
S.G. Tikhodeev, JETP Lett.  53  (3), 171 (1991).
\bibitem{Tikhodeev2}
S.G. Tikhodeev, Sol. St. Com. {\bf 78}, 339 (1991).
\bibitem{Kolesnikov3}
A.V. Kolesnikov, R. Lipperheide, A.P. Silin, U. Wille, Europhys. Lett. {\bf 43}, 331 (1998).
\bibitem{Andyushin5}
E.A. Andryushin, A.P. Silin, S.A. Vereshchagin, Phys. Low-Dim. Struct. {\bf 3/4}, 79 (2000).
\bibitem{Ratnikov3}
P.V. Ratnikov and A.P. Silin, Phys. Solid State {\bf 52} (8), 1763 (2010).
\bibitem{Abrikosov}
A.A. Abrikosov, J. Low Temp. Phys. {\bf 2}, 37 (1970).
\bibitem{Brazovsky}
S.A. Brazovskii, Sov. Phys. JETP {\bf 35} , 433 (1972).
\bibitem{Ando2}
T. Ando, J. Phys. Soc. Jpn. {\bf 66}, 1066 (1997).
\bibitem{Babichenko}
V.S. Babichenko, L.V. Keldysh, and A. P. Silin, Sov. Phys. Solid State {\bf 22} (4), 723 (1980).
\bibitem{Lozovik2}
Yu.E. Lozovik and V.I. Yudson, Phys. Lett. {\bf 56A}, 393 (1976).
\bibitem{Keldysh}
L.V. Keldysh, JETP Lett. {\bf 29} (11), 658 (1979).
\bibitem{Loudon}
R. Loudon, Am. J. Phys. {\bf 27}, 649 (1959).
\bibitem{Dalgarno}
A. Dalgarno and J.T. Lewis, Proc. R. Soc. A {\bf 233}, 70 (1955).
\bibitem{Miller}
D.A.B. Miller, D.C. Chemla, T.C. Damen et al., Phys. Rev. Lett. {\bf 53}, 2173 (1984).
\bibitem{Graf}
D. Graf, F. Molitor, K. Ensslin et al., Nano Lett. {\bf 7}, 238 (2007).
\bibitem{Ni}
Z.H. Ni, W. Chen, X.F. Fan et al., Phys. Rev. B {\bf 77}, 115416 (2008).
\bibitem{Ferrari}
A.C. Ferrari, J.C. Meyer, V. Scardaci et al., Phys. Rev. Lett. {\bf 97}, 187401 (2006).
\bibitem{Calizo}
I. Calizo, A.A. Balandin, W. Bao et al., Nano Lett. {\bf 7}, 2645 (2007).
\bibitem{Silin2}
A.P. Silin and S.A. Vereshchagin, Phys. Low-Dim. Struct. {\bf 9/10},
115 (2001).
\bibitem{Rashba1}
Yu.A. Bychkov and E.I. Rashba, JETP Lett. {\bf 39} (2), 78 (1984).
\bibitem{Rashba2}
Yu.A. Bychkov and E.I. Rashba, J. Phys. C: Solid State Phys. {\bf
17}, 6039 (1984).
\end{thebibliography}
\end{document}